\documentclass{ws-ijbc}
\usepackage{ws-rotating}     % used only when sideways tables/figures are used
\usepackage{graphicx}
\usepackage{epstopdf}
\begin{document}

\catchline{}{}{}{}{} % Publisher's Area please ignore

\markboth{A. Ishaq Ahamed and M. Lakshmanan}{Discontinuity Induced Hopf and Neimark-Sacker Bifurcations in a Memristive MLC Circuit}

\title{DISCONTINUITY INDUCED HOPF AND NEIMARK-SACKER BIFURCATIONS IN A MEMRISTIVE MURALI-LAKSHMANAN-CHUA CIRCUIT}

\author{A. ISHAQ AHAMED}

\address{Department of Physics, Jamal Mohamed College,\\ Tiruchirappalli-620020,India\\
ishaq1970@gmail.com}

\author{M. LAKSHMANAN}
\address{Centre for Nonlinear Dynamics,School of Physics,\\
Bharathidasan University,\\Tiruchirappalli-620024, India\\
lakshman@.gmail.com}

\maketitle

\begin{history}
\received{(to be inserted by publisher)}
\end{history}

\begin{abstract}
We report using Clarke's concept of generalised differential and a modification of Floquet theory to non-smooth oscillations, the occurrence of discontinuity induced Hopf bifurcations and Neimark-Sacker bifurcations leading to quasiperiodic attractors in a memristive Murali-Lakshmanan-Chua (memristive MLC) circuit. The above bifurcations arise because of the fact that a memristive MLC circuit is basically a nonsmooth system by virtue of having a memristive element as its nonlinearity. The switching and modulating properties of the memristor which we have considered endow the circuit with two discontinuity boundaries and multiple equilibrium points as well. As the Jacobian matrices about these equilibrium points are non-invertible, they are non-hyperbolic, some of these admit local bifurcations as well. Consequently when these equilibrium points are perturbed, they lose their stability giving rise to quasiperiodic orbits. The numerical simulations carried out by incorporating proper discontinuity mappings (DMs), such as the Poincar\'{e} discontinuity map (PDM) and zero time discontinuity map (ZDM), are found to agree well with experimental observations.

\end{abstract}

\keywords{memristive MLC circuit, non-hyperbolicity, switching manifolds, discontinuity induced bifurcations, Clarke's generalised differential, Hopf bifurcations and Neimark-Sacker bifurcations}

%\begin{multicols}{2}
\section{Introduction}

A memristor can be defined as any two-terminal device which exhibits a pinched hysteresis loop in the $(v-i)$ plane when driven by a bipolar periodic voltage or current excitation waveform, for any initial conditions \cite{ron14}. Its existence was first postulated by Leon O. Chua in the year 1971, purely from theoretical arguments \cite{chua71}.  Almost four decades later, Strutkov et al. of  Hewlett-Packard Laboratories designed the first approximate physical example of a memristor \cite{strut08} in a solid-state nano-scale system in which electronic and ionic transports are coupled under an external bias voltage. Immediately following this announcement, a theoretical study of memristive oscillators was made by \citet{itoh08}, which initiated a new field of research in memristive oscillators. 

The  $(v-i)$ characteristic of  memristors is inherently non-linear \cite{chua71}, because of which it is endowed with many peculiar features such as resistive switching at extremely high two-terminal device densities \cite{tour08}, pinched hysteresis characteristic and modulating ability \cite{icha11}. As a result of these, the memristor is considered as a desirable element in many  circuit applications. For example, memristor based chaotic circuits and their implementations were described in references \citep{muthu09a, muthu09b, muthu10a, muthu10b}. The existence of chaotic beats in a memristive Chua's circuit was reported in \citet{icha11} and hyperchoatic beats, hyperchaos and transient hyperchaos in a memristive Murali-Lakshmanan-Chua (MLC) circuit by \citet{icha13}. The coexistence of infinitely many stable periodic orbits and stable equilibrium points has been reported in memristive oscillators by \citet{mes10}. Chaos and its control in a four dimensional memristor based circuit using a twin-T Notch filter have been studied in \cite{Iu11}. Transient chaos has also been reported in a memristive canonical Chua's circuit by \citep{bao10a,bao10b}. The dynamical behaviour and stability analysis of a fractional-order memristor based Chua's circuit were investigated by \citet{pet10}. The nonlinear dynamics of a network of memristor oscillators was investigated by \citet{cor11}. Memristive chaotic circuits based on cellular nonlinear networks were studied by \citet{bus12}.
%------------------

In our earlier works on memristive circuits, we had identified the memristive Chua's circuit and memristive MLC circuit as non-smooth systems by virtue of the inclusion of a memristor as their nonlinear element. In particular, we identified that the primary cause of hyper-chaos, transient hyper-chaos and hyperchaotic beats that arose in memristive MLC circuit was the occurrence of \emph{grazing bifurcations}- a type of discontinuity induced bifurcation. But many other types of discontinuity bifurcations were found to arise in non-smooth systems, namely \emph{sliding bifurcations}, such as \emph{crossing-sliding}, \emph{switching-sliding}, \emph{grazing-sliding} and \emph{adding-sliding} bifurcations. In addition many other bifurcations such as \emph{impact oscillations} in a bouncing ball \citep{bab78,feigin94}, \emph{stick-slip} oscillations in mechanical systems \citep{gal97,gal01,gal98,bab78} were reported. Further \emph{discontinuity induced Hopf bifurcations} leading to \emph{limit cycle oscillations} were reported in non-smooth planar systems by \citet{leine06} and in the Chua's oscillator in two separate works by \citet{zhang12} and \citet{fu15}. Hence naturally a question arose as to the other types of bifurcations that may arise in the simple memristive MLC circuit. Our consequent works on the memristive MLC circuit led us to the identification of discontinuity induced \emph{Hopf} bifurcations and discontinuity induced \emph{secondary Hopf} or \emph{Neimark-Sacker} bifurcations leading to quasiperiodic oscillations and different types of \emph{sliding bifurcations}. In this paper we report the occurrence of discontinuity induced \emph{Hopf} bifurcations and discontinuity induced \emph{Neimark-Sackur} bifurcations in the memristive MLC circuit. The incidence of \emph{sliding} bifurcations in the memristive MLC circuit will be reported in subsequent articles.
%--------------------

The paper is organised as follows. In Sec. 2, we discuss the peculiar features of a the flux controlled memristor such as resistive switching property and modulating ability. In Sec. 3, the memristive MLC circuit, its circuit equations and their normalized forms using proper rescaling parameters are given. In Sec. 4, the memristive MLC circuit is considered as a piecewise smooth continuous system. Its switching manifolds are identified and the system equations are reformulated as a set of smooth ODEs.  The various equilibrium points admitted by this system are evaluated and their stability analysed. In Sec. 5, the memristive MLC circuit is considered for the  unforced case, and the discontinuity induced Hopf bifurcation exhibited by it under such a condition is studied by numerical simulations using Clarke's generalised differentials and by hardware experiments. In Sec. 6, the scaling up of the amplitude of the oscillations under the action of an external forcing is described through numerics. In Sec. 7, the discontinuity induced Neimark-Sackur bifurcation admitted by the same system under the action of an external forcing is studied by using a modification of the Floquet theory to non-smooth oscillations. In Sec. 8, a consolidation of the work is presented.

\section{Circuit Theoretic Properties: Resistive Switching and Memristive Modulation}
In this section we give a brief introduction to the circuit theoretic properties of the memristor such as \emph{resistive switching} and \emph{memristive modulation} and how they naturally lead to quasiperiodicity in the memristive MLC circuit.

Generally a memristor is a two terminal device whose flux-linkage $\phi$ and charge $q$ fall on some fixed curve in the $(q - \phi)$ plane and is represented mathematically as
\begin{equation}
g(\phi,q) = 0.
	\label{eqn:mem_rep}
\end{equation}
For a \emph{flux controlled} memristor this relation is expressed as a single-valued function of flux-linkage $\phi$. Consequently the constitutive relation becomes 
\begin{equation}
i(t) = W(\phi) v(t),
	\label{eqn:mem_rep_fc}
\end{equation}
where 
\begin{equation}
W(\phi) = \frac{dq(\phi)}{d\phi},
	\label{eqn:memductance}
\end{equation} 
defines the incremental \emph{memductance}. It is so called because it has the dimension of conductance, namely $\mho$ or Siemens. 

The time varying resistive property of the resistors reported by Chua in \cite{chua87} can be easily extended to memristors. From Eq. (\ref {eqn:memductance}) it is clear that the memductance $W(\phi)$ is a function of the flux-linkage $\phi$. But from Faraday's law, the flux linkage is defined as the time integral of the voltage, namely
\begin{equation}
\phi(t) = \int_{-\infty}^{t}v(\tau)d\tau
	\label{eqn:Faraday}
\end{equation}
Hence we can write
\begin{eqnarray}
\phi(t) & = & \int_{-\infty}^{t}v(\tau)d\tau \nonumber \\
				& = & \int_{-\infty}^{t_0}v(\tau)d\tau + \int_{t_0}^{t} v(\tau)d\tau, \: t\geq t_0 \nonumber \\
				& = & \phi(t_0) + \int_{t_0}^{t}v(\tau)d\tau,
	\label {eqn:mem_phi}
\end{eqnarray}
where $t_0$ is some chosen value of $t$.
If we assume $\phi(t_0) = 0$, then from the the above Eq. (\ref{eqn:mem_phi}) the value of the flux-linkage $\phi$ at any instant of time $t$ depends on the time integral of the voltage from some initial time $t_0$ to the instant of time $t$. Hence while a flux-controlled memristor behaves as an ordinary conductor at any given instant of time $t$, its total memductance depends on the complete past history of the memristor voltage. 

These variations of the memductance as a function of time cause the memristor to possess both a time varying resistive property as well as a memory nature. As a result of these, the memristor is endowed with some desirable characteristic features such as \textit{resistive switching property}, \textit{ modulating ability} etc., making it a desirable element in many circuit applications.

\subsection{Resistive Switching}
The resistive switching property of a memristor can be studied by using the analogy of an electrical switch. Let this switch be activated by sinusoidal variations of the flux($\phi$) having an angular frequency $\omega$ and a time period $\tau$. When the flux exceeds a predetermined level, say $B_p$, let the switch be in ON state and when the flux falls below this level $B_p$ let the switch be in OFF state. 
Mathematically this variation of the memductance can be represented as
\begin{equation}
W(t) = W(\phi(t)) = W(a) + W(b)\Phi(t),
	\label {eqn:W(t)}
\end{equation}
where $W(a)$ and $W(b)$ are the parametric memductance values of the memristor and $\Phi(t)$ represents the Fourier series for an anti-symmetric square wave of amplitude $\Phi_m$ and frequency $f_m$. 
\begin{eqnarray}
\Phi(t) & = & \frac{4\Phi_m}{\pi}\sum_{n=1}^{\infty}\dfrac{sin((2n-1)2\pi f_m t)}{(2n-1)} \nonumber \\
		& = &  \frac{4\Phi_m}{\pi} \left\lbrace \sin(2\pi f_m t) +\dfrac{1}{3}\sin(6\pi f_m t)+\dfrac{1}{5}sin(10\pi f_m t)+ ........ \right\rbrace.
	\label {eqn:f(t)}
\end{eqnarray}

\begin{figure}[htbp]
	\centering
	\resizebox{0.95\textwidth}{!}
		{\includegraphics{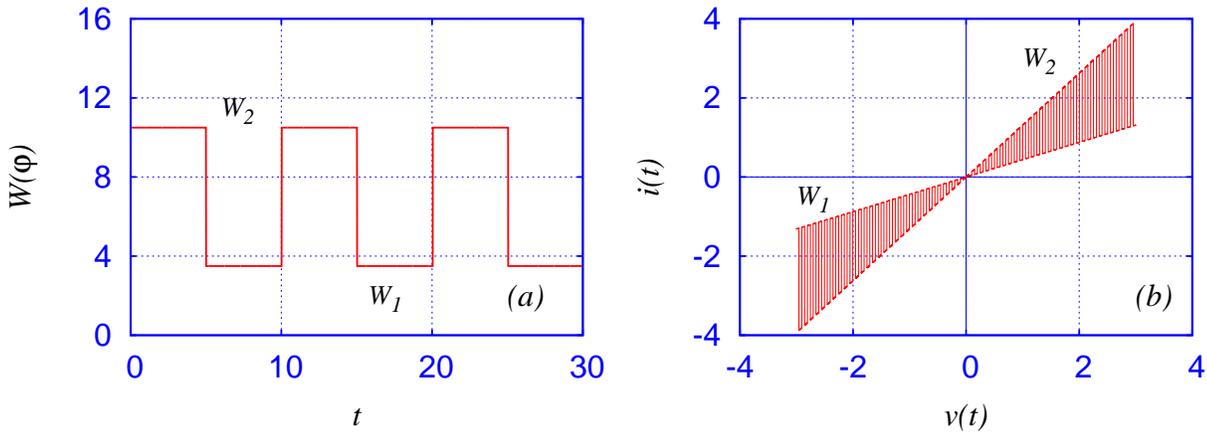}}
	\caption[Memristor Switching Characteristic] {Memristor Resistive Switching Characteristic:(a) Variation of memductance $W(\phi(t))$ as a function of time $t$ and (b) variations of $(v-i)$ characteristic due to memristor switching.}
	\label{fig:mem_chac1}
\end{figure}

The variations in parametric memductance values causes the resultant memductance $W(\phi)$ of the memristor to vary as a function of time $t$, with a low OFF state value of $W_1 = W(a)$ and high ON state value of $W_2 = W(a) + W(b)\left( \dfrac{4\Phi_m}{\pi} \sum\limits_{n=1}^{\infty}\dfrac{sin((2n-1)2\pi f_m t)}{(2n-1)}\right)$. These variations take place at a constant frequency $f_m $ and define the resistive switching characteristic of the memristor with time $t$, as shown in Fig. \ref{fig:mem_chac1}(a).

From Fig. \ref{fig:mem_chac1}(a) we find that the memristor behaves as a practical resistive switching element as postulated by Chua in \cite{chua11}. The memristor has a very low memductance $W_1$ during the time $0 < t \leq t_1$ and a very large memductance $W_2$ during the time $t_1 < t \leq T$. During the time the memductance is low, the memristor switch is said to be open $(S = 0)$, while during the time the memductance is high, the switch is said to be closed $(S=1)$. After a period of time $T$, the switch repeats its operation, that is, the memristor takes on states $S=0$ and $S=1$ alternately for $nT < t \leq (n+1)T$, where $n$ is any integer $>1$. As the switching time $T$ of the memristor is inherently very small when compared to mechanical switches, it is suitable for digital applications.

The $(v-i)$ characteristic is obtained by plotting the current $i(t)$ from Eq. 
(\ref{eqn:mem_rep_fc}), as a function of the memristor voltage $v(t)$.  This 
characteristic is a set of two straight lines with slopes equal to the 
memductance value $W_1$ when the memristor is in an OFF state and $W_2$ when 
in the ON state respectively. Hence the characteristic in the $(v-i)$ plane 
alternates between these two straight lines as it constantly switches from the 
OFF state to ON state and vice-versa as shown in Fig. \ref{fig:mem_chac1}(b).

\subsection{Memristor Modulation}
\begin{figure}[htbp]
	\centering
	\resizebox{0.95\textwidth}{!}
		{\includegraphics{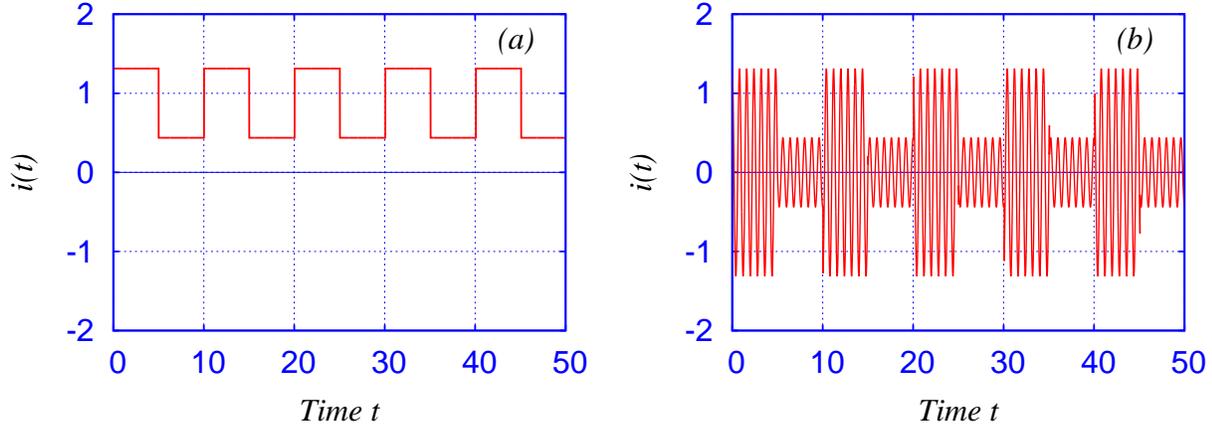}}
	\caption[Memristor Modulation] {Memristor Modulation: (a) Variation of the current $i(t)$ across the memristor when a constant potential $v_{dc}$ is applied and (b) the resultant current $i(t)$ when a periodic voltage $v(t) = v_0\cos(2\pi f_e t)$ is applied. It is clear that the superposed current $i(t)$ in the second case is a modulated current with three frequency components, namely, $f_e$ the frequency of the external force, and two components $f_e \pm (2n-1)f_m$, $n = 1,2,3,...$ arising due to the interaction of the external force with the memristor oscillations of frequency $f_m$.}
	\label{fig:mem_mod_chac1}
\end{figure}

The current $i(t)$ across the memristor, as is seen from Eq. 
(\ref{eqn:mem_rep_fc}), varies not only with the voltage $v(t)$ applied across 
it, but also with the memductance $W(\phi(t))$. Let a dc voltage be applied to 
the memristor such that
\begin{equation}
v(t) = v_{dc}.
	\label {eqn:E_dc}
\end{equation}
Then the current across the memristor should possess a time varying nature due to the resistive switching property. This is actually the case as is shown in Fig. \ref{fig:mem_mod_chac1}(a).
Let the voltage $v(t)$ driving the memristor be a periodic voltage, say for example a sinusoid with a frequency $f_e$, that is
\begin{equation}
v(t) = v_0 cos(2\pi f_e t).
	\label {eqn:e(t)}
\end{equation}
Then on substituting Eqs. (\ref {eqn:W(t)}) and (\ref {eqn:e(t)}) into Eq. (\ref {eqn:mem_rep_fc}), the current across the memristor is
\begin{eqnarray}
i(t) & = & \left \lbrace W(a) + W(b)\frac{4\Phi_m}{\pi}\sum\limits_{n=1}^{\infty}\frac{sin((2n-1)2\pi f_m t)}{(2n-1)}\right \rbrace v_0 cos(2\pi f_e t) \nonumber \\ \nonumber \\
     & = & W(a)v_0 cos(2\pi f_e t) + \nonumber \\  \nonumber \\
     &  & W(b)\dfrac{4\Phi_m v_0}{\pi} \sum\limits_{n=1}^{\infty} \left \lbrace \dfrac{\sin((2n-1)2\pi f_m t)}{(2n-1)} \right\rbrace v_0 cos(2\pi f_e t) \nonumber\\  \nonumber
     & = & W(a)v_0 cos(2\pi f_e t) + \nonumber \\ \nonumber \\
     &  & \left( \dfrac{4W(b)\Phi_m v_0}{2\pi}\right) 
     \sum\limits_{n=1}^{\infty}\left\lbrace \dfrac{sin(2\pi (f_e+(2n-1)f_m)t)}{(2n-1)}\right\rbrace  + \nonumber\\ \nonumber \\
     &  & \left(  \dfrac{4W(b)\Phi_m v_0}{2\pi}\right) \sum\limits_{n=1}^{\infty}\left\lbrace  \dfrac{sin(2\pi (f_e-(2n-1)f_m)t)}{(2n-1)}\right\rbrace .     
	\label{eqn:W(t)_modulation}
\end{eqnarray}
Thus we find that in addition to the frequency $f_e$ of the driving voltage, the memristor causes the circuit to generate new currents which are sinusoids with frequencies $(f_e \pm (2n-1)f_m), n = 1,2,3,.....$. The superposed current of all the sinusoids is shown in Fig. \ref{fig:mem_mod_chac1}(b). 
\begin{figure}[htbp]
	\centering
	\resizebox{0.95\textwidth}{!}
		{\includegraphics{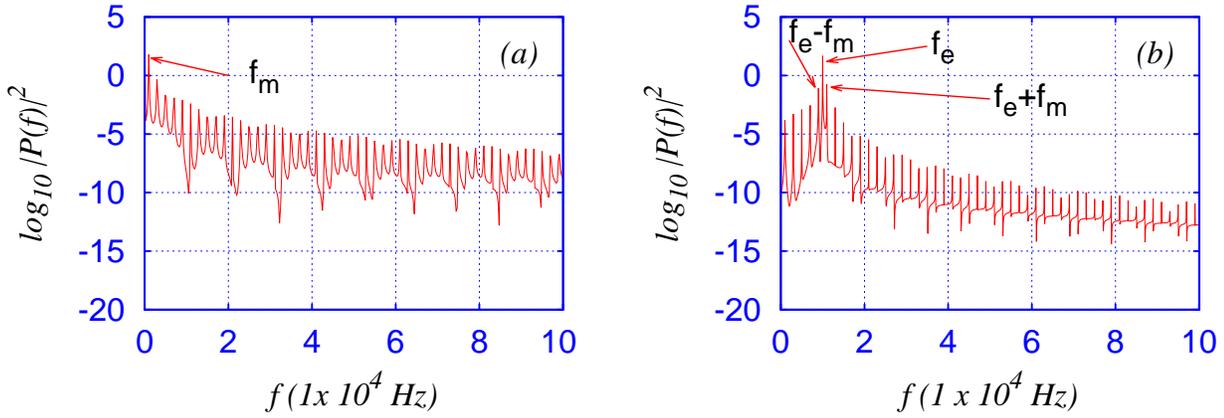}}
	\caption[Power Spectra of Memristor Modulation] {Power Spectra of Memristor Modulation : Power spectrum of the current $i(t)$ across the memristor (a) showing a single peak at $f_m$ when a constant voltage $v_{dc}$ is applied and (b) showing peaks at $f_e$ and at $f_e \pm (2n-1)f_m$ when a sinusoidal voltage $v(t) = v_0\cos(2\pi f_e t)$ is applied. The numerical values are: $f_m = 1000 Hz$, $f_e = 10000 Hz$, $(f_e - f_m) = 9000 Hz $ and $(f_e + f_m)=  11000 Hz $.}
	\label{fig:mem_mod_spectrum}
\end{figure}

The frequencies of currents across the memristor when it is driven by a constant voltage and a sinusoidal voltage can be found out from power spectral analysis as shown in Fig. \ref{fig:mem_mod_spectrum}. The power spectrum of the current across the memristor when it is driven by a constant voltage $v_{dc}$ shows a single peak at $f_m$ and its harmonics. Similarly the power spectrum of the current across the memristor when driven by a sinusoidal voltage shows three prominent peaks, one at the driving frequency $f_e$, and the other two at $(f_e - f_m)$ and at $(f_e + f_m)$ , in addition to the various subharmonic and super harmonic frequency components.

It is evident from the above discussion that had the memristor possessed a purely time invariant nature, that is, if $W(b) = 0$, it would cause the circuit to generate just a single sinusoid response with the original driving frequency $f_e$ only.

This ability of the memristor to introduce additional frequencies and harmonics in the circuit whenever it is driven by sinusoidal or any other periodic signal causes limit cycle oscillations in memristive circuits to take on a quasiperiodic nature as will be seen in the following sections.

\section{Memristive Murali-Lakshmanan-Chua Circuit} 

\begin{figure}[h]
	\begin{center}
		\psfig{file=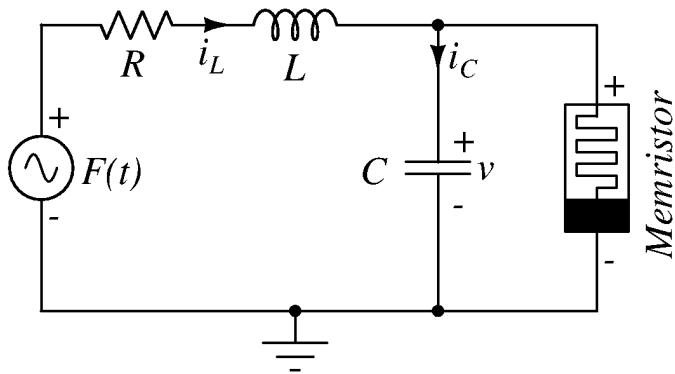,width=3.5in} 
	\end{center}
	\caption[Memristive MLC Circuit] {The memristive MLC circuit}
	\label{fig:mmlc_cir}
\end{figure}

\begin{figure}[h]
	\begin{center}
	\psfig{file=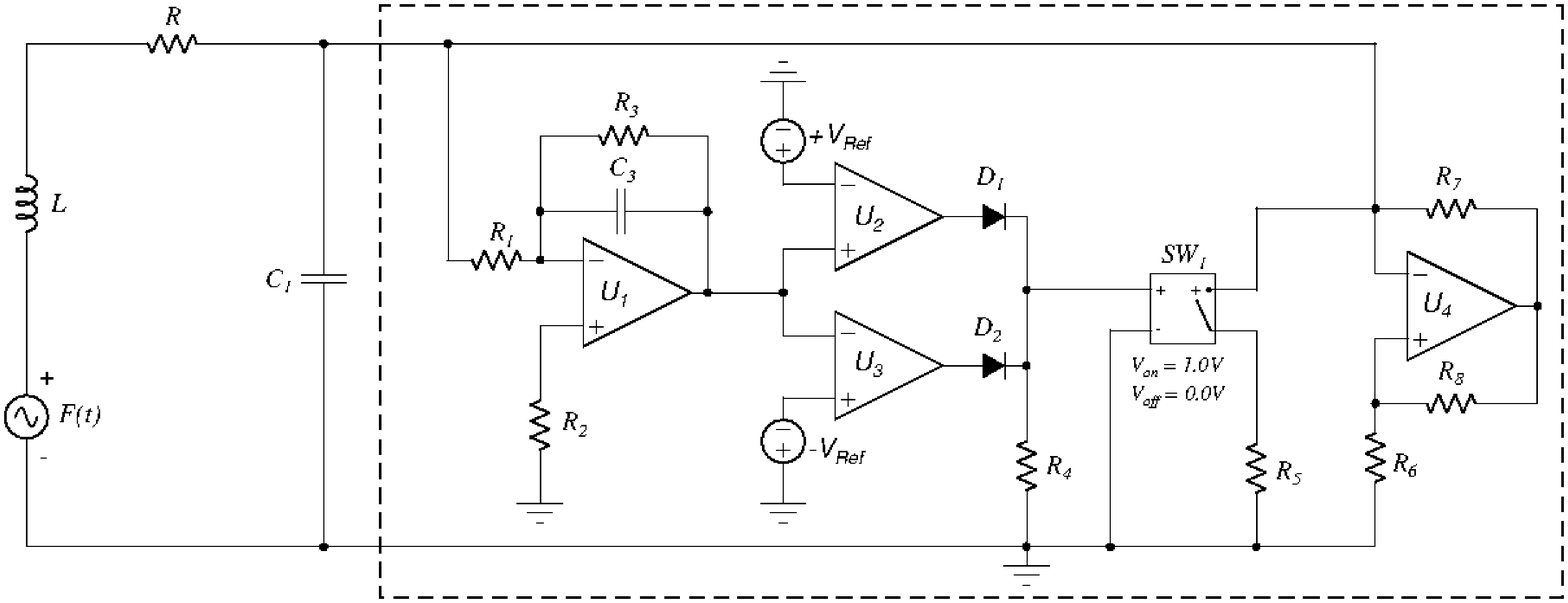,width=6.5in} 
	\end{center}
		\caption[Multisim Prototype of the Memristive MLC Circuit] {A Multisim Prototype Model of a memristive MLC circuit. The memristor part is shown by the dashed outline. The parameter values of the circuit  are fixed as $ L = 21 mH$, $R = 900 \Omega$, $C_1 = 10.5nF$. The frequency of the external sinusoidal forcing is fixed as $\nu_{ext} = 8.288 kHz$ and the amplitude is fixed as $F = 770 mV_{pp} $ ( peak-to-peak voltage). }
		\label{fig:prototype}
\end{figure}

A memristive MLC circuit, constructed by replacing the Chua's diode in the classical Murali-Lakshmanan-Chua circuit with an analog model of an active flux controlled memristor, proposed by \cite{icha11} as its non-linear element was reported earlier in 2013, refer \citet{icha13}. The schematic circuit is shown in Fig. \ref{fig:mmlc_cir}, while the actual analog realization based on the prototype model for the memristor is shown in Fig. \ref{fig:prototype}.
%----------
Applying Kirchoff's laws, the circuit equations can be written as a set of autonomous ordinary differential equations (ODEs) for the flux $\phi(t)$, voltage $v(t)$, current $i(t)$ and the time $p$ in the extended coordinate system as
\begin{eqnarray}
\frac{d\phi}{dt}  & = & v, \nonumber \\
C\frac{dv}{dt}  & = & i - W(\phi)v,  \nonumber \\ 
L \frac{di}{dt}  & = &  -v -Ri +F\sin( \Omega p),\nonumber \\
\frac{dp}{dt}  & = & 1.
	\label{eqn:mlc_cir}
\end{eqnarray}
Here $W(\phi)$ is the memductance of the memristor and is as defined in \cite{itoh08},
\begin{equation}
W(\phi) = \frac{dq(\phi)}{d\phi} = \left\{
					\begin{array}{ll}
					G_{a_1}, ~~~ | \phi  | > 1  \\
					G_{a_2}, ~~~ | \phi  | \leq 1,	
					\end{array}
				\right.
	\label{eqn:W}
\end{equation}
where $G_{a_1}$ and $G_{a_2}$ are the slopes of the outer and inner segments of the characteristic curve of the memristor respectively. We can rewrite Eqs. (\ref{eqn:mlc_cir}) in the normalized form as
\begin{eqnarray}
\dot{x}_1  & = & x_2, \nonumber \\
\dot{x}_2  & = & x_3-W(x_1)x_2, \nonumber \\ 
\dot{x}_3  & = & -\beta(x_2+x_3) + f \sin(\omega x_4),\nonumber \\
\dot{x}_4  & = &  1.
\label{eqn:mlc_nor}
\end{eqnarray}
Here dot stands for differentiation with respect to the normalized time $\tau$ (see below) and $W(x_1)$ is the normalized value of the memductance of the memristor, given as
\begin{equation}
W(x_1) = \frac{dq(x_1)}{dx_1} = \left\{
		\begin{array}{ll}
		a_1, ~~~ | x_1  | > 1 \\
		a_2, ~~~ | x_1  | \leq  1
		\end{array}
	\right.
	\label{eqn:W_nor}
\end{equation}
where $ a_1 = G_{a_1}/G $ and $a_2 = G_{a_2}/G$  are the normalized values of $G_{a_1} $ and $G_{a_2} $  mentioned earlier and are negative. The rescaling parameters used for the normalization are
\begin{eqnarray}
x_1 = \frac{G\phi}{C},x_2 = v, x_3 = \frac{i}{G}, x_4 = \frac{Gp}{C}, G = \frac{1}{R},\beta = \frac{C}{LG^2}, \\ \nonumber
\omega = \frac{\Omega C}{G} = \frac{2\pi \nu C}{G},\tau = \frac{Gt}{C},f = F\beta.
	\label{eqn:rescale} 
\end{eqnarray}

In our earlier work on this memristive MLC circuit, see \citet{icha13}, we reported that the addition of the memristor as the nonlinear element converts the system into a piecewise-smooth continuous flow having two discontinuous boundaries, admitting \emph{Grazing bifurcations}, a type of discontinuity induced bifurcation (DIB). These Grazing bifurcations were identified as the cause for the occurrence of hyperchaos, hyperchaotic beats and transient hyperchaos in this memristive MLC system. 

As stated in Sec. 1, piecewise-continuous systems possess rich dynamics, and are known to exhibit many types of non-smooth bifurcations, such as \emph{sliding} bifurcations, \emph{stick-slip} oscillations, \emph{impact} oscillations, etc. As the memristive MLC circuit is a piecewise-continuous system, we naturally realized that it can be no exception and that it should exhibit many other hitherto unknown \emph{discontinuity induced} bifurcations. Hence we made further investigations on this system by applying non-smooth bifurcation theory which has been well developed and applied to mechanical systems. These investigations led us to the observation of \emph{sliding} bifurcations and \emph{Discontinuity Induced Hopf} and \emph{Neimark-Sacker} bifurcations thus establishing the robustness and versatility of this simple low dimensional circuit.

\section{Memristive MLC Circuit as a Non-smooth System}

The memristive MLC circuit is a piecewise-smooth continuous system by virtue of the discontinuous nature of its nonlinearity, namely the memristor. Referring to the memductance characteristic, we find that the memristor switches states at $x_1 = +1$ and at $x_1 = -1$ either from a more conductive ON state to a less conductive OFF state or vice versa. These switching states of the memristor give rise to two discontinuity boundaries or switching manifolds, $\Sigma_{1,2}$ and $\Sigma_{2,3}$ which are symmetric about the origin and are defined by the zero sets of the smooth functions $H_i(\mathbf{x},\mu) = C^T\mathbf{x}$, where $C^T = [1,0,0,0]$ and $\mathbf{x}= [x_1,x_2,x_3,x_4]$, for $i=1,2$. Hence $H_1(\mathbf{x}, \mu) = (x_1-x_1^\ast)$, $x_1^\ast = -1$ and $H_2(\mathbf{x}, \mu) = (x_1-x_1^\ast)$, $x_1^\ast = +1$, respectively. Consequently the phase space $\mathcal{D}$ can be divided into three subspaces $S_1$, $S_2$ and $S_3$ due to the presence of the two switching manifolds. The memristive MLC circuit can now be rewritten as a set of smooth ODEs 
\begin{equation}
\dot{x}(t) = 
	\left\lbrace	\begin{array}{lcccl}
		F_2(\mathbf{x},\mu),& H_1(\mathbf{x}, \mu) >  0& \& & H_2(\mathbf{x}, \mu) < 0,& \mathbf{x} \in S_2,  \\
				   &					&    &                 &           \\
		F_{1,3}(\mathbf{x},\mu), &  H_1(\mathbf{x}, \mu)< 0 &\& &H_2(\mathbf{x},\mu)> 0, &\mathbf{x} \in S_{1,3},
		\end{array}
	\right.
	\label{eqn:smooth_odes}
\end{equation}
where $\mu$ denotes the parameter dependence of the vector fields and the scalar functions. The vector fields $F_i$'s are
\begin{equation}
 F_i(\mathbf{x},\mu) =  \left (	\begin{array}{c}
					x_2			\\
				-a_i x_2+x_3	\\
				-\beta x_2 -\beta x_3 +f sin(\omega x_4)	\\	
				1 
				\end{array}
		\right ), \mathrm{i\;=\;1,2,3}
\label{eqn:vect_field}
\end{equation}
where we have $a_1 = a_3$. 

The discontinuity boundaries $\Sigma_{1,2}$ and $\Sigma_{2,3}$ are not uniformly discontinuous. This means that the degree of smoothness of the system in some domain $\mathcal{D}$ of the boundary is not the same for all points $x \in \Sigma_{ij}\cap \mathcal{D}$. This causes the memristive MLC circuit to behave as a non-smooth system having a degree of smoothness of either {\textit{one}} or {\textit{two}}. In such a case it will behave either as a \emph{Filippov system} or as a \emph{piecewise-smooth continuous flow} respectively, refer Appendix A. Hence the memristive MLC circuit is found to admit different types of \emph{Discontinuity Induced Bifurcations} (DIB's) such as \emph{Boundary Equilibrium Bifurcations} (BEB's), \emph{Crossing-Sliding} bifurcations, \emph{Grazing-Sliding} bifurcations, \emph{Switching-Sliding} bifurcations and pure \emph{Grazing} bifurcations. In this paper we restrict our description to the boundary equilibrium bifurcations alone that are admitted by this system. \emph{Boundary Equilibrium Bifurcations} are the simplest type of DIB's that occur in piecewise-smooth continuous systems \citep{dib02a,leine02,leine04}. They occur if for a particular critical value of the parameter the equilibrium point lies precisely on the discontinuity boundary $\Sigma$. In this paper we will show that the replacement of the Chua's diode by an active flux controlled memristor in the MLC circuit results in the system possessing boundary equilibrium points and exhibiting discontinuity induced Hopf and Neimark-Sacker bifurcations about these points. We will analyse the equilibrium points of the circuit, their stability and the dynamics arising from these DIBs in the following sections. We also note here that any other nonsmooth system having a degree of smoothness two or greater than two can be shown to exhibit similar kind of boundary equilibrium points and associated bifurcations.

\subsection{Equilibrium Points and their Stability}
In the absence of the driving force, that is if $f=0$, the memristive MLC circuit can be considered as a three-dimensional autonomous system with vector fields given by
\begin{equation}
 F_i(\mathbf{x},\mu) =  \left (	
 					\begin{array}{c}
							x_2			 	\\
						-a_i x_2+x_3  		\\
						-\beta x_2 -\beta x_3	\\	
					\end{array}
		\right ), \mathrm{i\;=\;1,2,3}.
\label{eqn:vect_field3d}
\end{equation}

\begin{figure}[htbp]
 	\begin{center}
 			\psfig{file=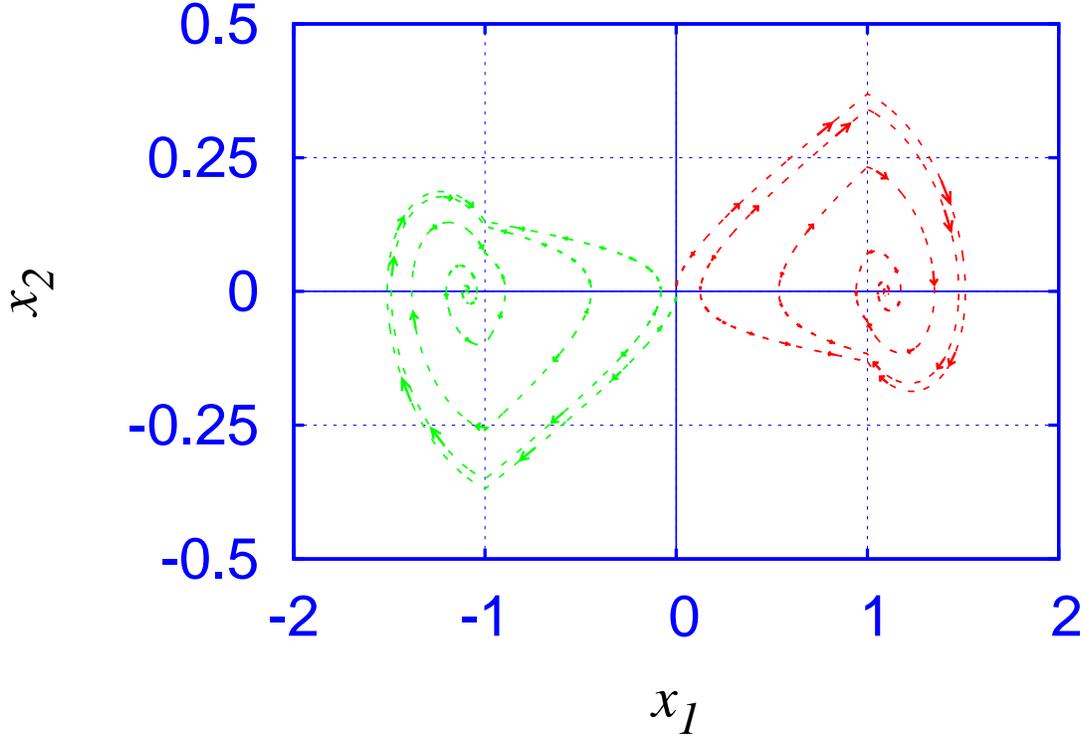,width=6.5in} 
 	\end{center}
 	\caption[Equilibrium Points]{Figure showing the equilibrium points $E_{\pm}$ in the subspaces $S_1$ and $S_3$ for the parameter value above $\beta_c = 0.8250$. The initial conditions are  $x_1 = 0.0$, $x_2 = 0.01$, $x_3 = 0.01$ for the fixed point $E_{+}$ in the subspace $S_3$ and 
 	$x_1 = 0.0$, $x_2 = -0.01$, $x_3 = -0.01$ for the fixed point $E_{-}$ in the subspace $S_1$.}
 	\label{fig:mmlc_limitcycle}
 \end{figure}

This three dimensional autonomous system has a trivial equilibrium point $E_0$,
two {\textit{admissible equilibrium}} points $E_{\pm}$ and two {\textit{boundary equilibrium}} points $E_{B\pm}$ (see Appendix A for details). The multiplicity of equilibrium points arises because of the non-smooth nature of the nonlinear function, namely $W(x_1)$ given in Eq. (\ref{eqn:W_nor}). The trivial equilibrium point is given as
\begin{equation}
E_0 = \{(x_1,x_2,x_3)|x_1=x_2=x_3=0\}
\end{equation}
The two admissible equilibria $E_{\pm}$ are
\begin{equation}
E_{\pm} = \{(x_1,x_2,x_3)|x_2=x_3=0,x_1^*= \textrm{constant and not equal to } \pm 1 \}
\end{equation}
At the admissible equilibrium points $E_{\pm}$, the conditions given by Eqs. (\ref{eqn:pw_admissible1}) and (\ref{eqn:pw_admissible2}) are satisfied, that is 
\begin{eqnarray}
F_1(\mathbf{x},\mu) = 0,\\ \nonumber
H_1(\mathbf{x},\mu) \text{:} = \lambda_1 < 0, \nonumber 
	\label{eqn:mmlc_admissible1}
\end{eqnarray}
or
\begin{eqnarray}
F_3(\mathbf{x},\mu) = 0, \\ \nonumber
H_2(\mathbf{x},\mu) := \lambda_2 > 0. 
	\label{eqn:mmlc_admissible2}
\end{eqnarray}
As the vector fields $F_1(\mathbf{x},\mu)$ and $F_3(\mathbf{x},\mu)$ are symmetric about the origin, that is $F_1(\mathbf{x},\mu) = F_3(-\mathbf{x},\mu)$, the admissible equilibria $E_{\pm}$ are also placed symmetric about the origin in the subspaces $S_1$ and $S_3$. These are shown in Fig. \ref{fig:mmlc_limitcycle} for a certain choice of parametric values. For plotting the stable focus $E_+$ in $S_3$ we assumed the initial conditions as $x_1 = 0.0$, $x_2 = 0.01$, $x_3 = 0.01$ while for the stable focus $E_-$ in the subspace $S_1$ the initial conditions are $x_1 = 0.0$, $x_2 = -0.01$, $x_3 = -0.01$.
 
At the boundary equilibrium points $E_{B\pm}$,
\begin{equation}
E_{B\pm} = \{(x_1,x_2,x_3)|x_2=x_3=0,\hat{x}_1= \pm 1 \}
\end{equation}
the conditions given by Eq. (\ref{eqn:pw_BEB1}) are satisfied, namely
\begin{equation}
\begin{array}{ccc}
F_{1,2}(\hat{\mathbf{x}},\mu)  = 0, & \qquad H_1(\hat{\mathbf{x}},\mu) = 0 & \qquad \textit{or}\\
F_{2,3}(\hat{\mathbf{x}},\mu)  = 0, & \qquad H_2(\hat{\mathbf{x}},\mu) = 0 &
\end{array}
		\label{eqn:mmlc_BEB1}
\end{equation}
These boundary equilibria lie on the switching manifolds $\Sigma_{1,2}$ and $\Sigma_{2,3}$.

The definitive conditions for the existence of equilibria given above however do not tell us about the stability of these equilibrium states. Therefore we construct the Jacobian matrices $N_i, \, i = 1,2,3$ and evaluate their eigenvalues at these points to analyse their stability,
\begin{equation}
N_i =   \left ( \begin{array}{ccc}
				0      & 1    	&  0  		 \\
				0      &-a_i    &  1 		 \\
				0      &-\beta  &  -\beta 	\\				 
				\end{array}
		\right),  \text{i\;=\;1,2,3}.
	\label{eqn:Jac}
\end{equation}
The characteristic equation associated with the system $N_i$ in these equilibrium states is
\begin{equation}
\lambda^3 + p_2\lambda^2 + p_1 \lambda = 0,
	\label{eqn:chac}
\end{equation}
where {\it{$\lambda$}}'s are the eigenvalues that characterize the equilibrium states and $\it{p_i}$'s are the coefficients, given as $p_1 = \beta(1+a_i)$ and $p_2 = (\beta + a_i)$. The eigenvalues are 
\begin{equation}
\lambda_1 = 0, \,\lambda_{2,3} = \frac{-(\beta+a_i)}{2} \pm \frac{\sqrt{(\beta - a_i)^2-4 \beta}}{2}.
	\label{eqn:eigen}
\end{equation} 
where $ i = 1,2,3$. Depending on the eigenvalues, the nature of the equilibrium states differ. 
\begin{enumerate}

\item
When $(\beta - a_i)^2 = 4 \beta$, the equilibrium state will be a stable/unstable star depending on whether $(\beta+a_i)$ is positive or not.

\item
When $(\beta - a_i)^2 > 4 \beta$, the equilibrium state will be a saddle.

\item
When $(\beta - a_i)^2 < 4 \beta$, the equilibrium state will be a stable/unstable focus.
\end{enumerate}
For the third case, the circuit admits self oscillations with natural frequency varying in the range $\sqrt{\left [(\beta - a_1)^2-4 \beta \right ]} / 2 < \omega_o < \sqrt{\left [(\beta - a_2)^2-4 \beta \right ]}/2$. It is at this range of frequency that the memristor switching also occurs. 

\subsection{Boundary Equilibrium Bifurcations (BEB)}
The boundary equilibrium bifurcations are analogous to the border-collision bifurcations in maps. For a piecewise smooth continuous flow, two types of boundary equilibrium bifurcations have been observed, namely \emph{persistence} and \emph{non-smooth fold}, see \cite{dib08} and the references mentioned therein. The conditions for the existence of these bifurcations are given in Appendix - A. Applying them, we find that the memristive MLC circuit does not admit any {\textit{boundary equilibrium}} bifurcations. This can be explained as due to the non-invertibility of the Jacobian matrices.

Let us linearise the memristive MLC circuit described by the Eqs. (\ref{eqn:smooth_odes}) and (\ref{eqn:vect_field}) about the boundary equilibrium points $E_{B\pm}$ as referred to in Eqs (\ref{eqn:beba1}) and (\ref{eqn:beba2}) in Appendix A, that is
\begin{eqnarray}
N_1\; \mathbf{x^-} +M_1\;\mu & = & 0, \nonumber \\
C^T \;\mathbf{x^-} +D\;\mu & = & \lambda^-,
	\label{eqn:sec_beba1}
\end{eqnarray}
and
\begin{eqnarray}
N_2\;\mathbf{x^+} +M_2 \;\mu & = & N_1\mathbf{x}^+ M_1\mu + E\lambda^+ = 0, \nonumber \\
C^T\mathbf{x^+} +D\;\mu & = & \lambda^+,
		\label{eqn:sec_beba2}
\end{eqnarray}
where $N_1 = F_{1,x}, M_1 = F_{1,\mu}$, $C^T=H_x$, $D = H_{\mu}$ and $\mathbf{x^+}(\mu)$ and $\mathbf{x^-}(\mu)$ are the branches of the equilibria, before and after the crossing of the discontinuity boundaries at $\mu = \mu^\ast$ in the subspaces $S_1$ and $S_2$ of the vector fields $F_1(\mathbf{x},\mu))$ and $F_2(\mathbf{x},\mu)$ respectively. Then
\begin{equation}
		\left ( \begin{array}{ccc}
				0      & 1    	&  0  		 \\
				0      &-a_1    &  1 		 \\
				0      &-\beta  &  -\beta 	\\				 
				\end{array}
		\right )\mathbf{x^-} +
		\left ( \begin{array}{c}
						0      \\
						1      \\
						1      \\				 
				\end{array}
		\right) \mu = 0 \nonumber
\end{equation}
\begin{equation}
		\left ( \begin{array}{c}
						1		\\
						0		\\
						0		\\
				\end{array}		
		\right )\mathbf{x^-} +D\mu = \lambda^-,
		\label{eqn:sec_mmlcbec1}	
\end{equation}
and
\begin{equation}
		\left ( \begin{array}{ccc}
				0      & 1    	&  0  		 \\
				0      &-a_1    &  1 		 \\
				0      &-\beta  &  -\beta 	\\				 
				\end{array}
		\right )\mathbf{x^+} +
		\left ( \begin{array}{c}
						0      \\
						1      \\
						1      \\				 
				\end{array}
		\right) \mu +
		\left ( \begin{array}{c}
								0      \\
								(a_1-a_2)      \\
								0      \\				 
						\end{array}
				\right ) \lambda^+ = 0, \nonumber
\end{equation}
\begin{equation}
		\left ( \begin{array}{c}
						1		\\
						0		\\
						0		\\
				\end{array}		
		\right ) \mathbf{x^+} + D\mu = \lambda ^+.
	\label{eqn:sec_mmlcbeb2}		
\end{equation}
where we have taken $N_2-N_1 = E$, with 
$E = \left( 
		\begin{array}{c}
			0     		   \\
			(a_1-a_2)      \\
			0      			\\				 
		\end{array}
	\right ).
$
 Thus we find that as Eqs. (\ref{eqn:sec_mmlcbec1}) and (\ref{eqn:sec_mmlcbeb2}) are satisfied, the system possess two admissible equilibrium points $E_{\pm}$. \\

However as the determinant of the Jacobian matrix $N_1$ is zero, that is as   $det(N_1)=0$, we find that the Jacobian matrix is not invertible. Consequently the conditions, refer Eqs. (\ref{eqn:beba1}), (\ref{eqn:beba2}), (\ref{eqn:persistence}) and (\ref{eqn:nonsmooth_fold}) for the occurrence of {\textit{boundary equilibrium}} bifurcations are not satisfied. Therefore based on this we conclude that {\textit{boundary equilibrium}} bifurcations are not admitted by the memristive MLC circuit.

The non-invertibility of the Jacobian matrices, however suggests that one or more of the many equilibrium points may be {\textit{non-hyperbolic}} and may admit local bifurcations with respect to one of the vector fields, either $F_1$, $F_2$ or $F_3$. This is exactly what happens in the memristive MLC circuit, for we find that  {\textit{discontinuity induced Hopf }} bifurcations and {\textit{discontinuity induced Neimark-Sacker}} bifurcations occur leading to complex dynamics. As noted earlier, this system can also exhibit even more complicated bifurcations like sliding bifurcations, apart from the grazing bifurcations reported in our earlier work, \cite{icha13}. These more sophisticated bifurcations will be reported separately.

\section{Discontinuity Induced Hopf Bifurcation}
The existence of invariant sets, other than equilibrium points, and their creation and destruction in a boundary equilibrium bifurcation has been reported in \cite{leine00},\cite{leine04} and \cite{leine02}. Here in the memristive MLC circuit, as boundary equilibrium bifurcations are not observed, we find that perturbing the boundary equilibrium point about either of the switching manifolds $\Sigma_{1,2}$ or $\Sigma_{2,3}$ results in the trajectory forming a {\textit{quasi-periodic}} attractor. The birth of the quasi-periodic attractor can be attributed to the occurrence of a {\textit{discontinuity induced Hopf}} bifurcation at the switching manifolds and {\textit{memristor modulation}}. Such discontinuity induced Hopf bifurcations leading to {\textit{limit cycle}} motion in planar non-smooth systems were reported for the first time by \citet{leine06} and in the Chua's oscillator in two separate works by \citet{zhang12} and \citet{fu15}. For identifying this we apply the extension of the concept of Clarke's generalised differential to Jacobian matrices in order to evaluate mathematically the correct eigenvalues, and have also constructed proper discontinuity mappings such as the \emph{Zero Time Discontinuity Map} (ZDM) and \emph{Poincar\'{e} Discontinuty Map} (PDM). The details of these discontinuity mapping corrections have been elaborated in our earlier work, see \cite{icha13}, and are therefore not provided here for the sake of brevity. 
 
Let us assume the normalised parameter values of the memristor MLC circuit to be $a_{1,3} = -0.55$, $a_2 = -1.02$. Let $\beta$ be the control parameter. For the value $\beta = 0.825$, the eigenvalues of the Jacobian matrix $N_2$ for the subspace $S_2$ in the central region are $\lambda_{\mathbf{2},1} = 0$, $\lambda_{\mathbf{2},2} = 0.258765$ and $\lambda_{\mathbf{2},3} =  -0.0637645$, while the eigenvalues of the Jacobian matrices $N_{1,3}$ for the subspaces in the outer two regions, namely $S_1$ and $S_3$ are $\lambda_{\mathbf{(1,3)}1} = 0$, $\lambda_{\mathbf{(1,3)}2,3} = -0.1375 \;\pm\; i\; 0.5935861$. Looking at the eigenvalues of the Jacobian matrices we find that the trivial equilibrium point $E_0$ in the subspace $S_2$ is an unstable {\textit{saddle}} while the admissible points $E_{\pm}$ in the subspaces $S_{1}$ and $S_3$ are {\textit{stable foci}} for these parametric values, refer Fig. \ref{fig:mmlc_limitcycle}.

Left to themselves, the stability of these equilibrium points will remain unaffected. But when the trajectories cross the switching manifolds $\Sigma_{1,2}$ and $\Sigma_{2,3}$ their behaviour is affected by the vector fields $F_1$ and $F_2$ on both sides of the switching boundaries. The combined effect of the two vector fields is so as to cause the birth of periodic orbits.
\begin{figure}[htbp]
	\begin{center}
		\psfig{file=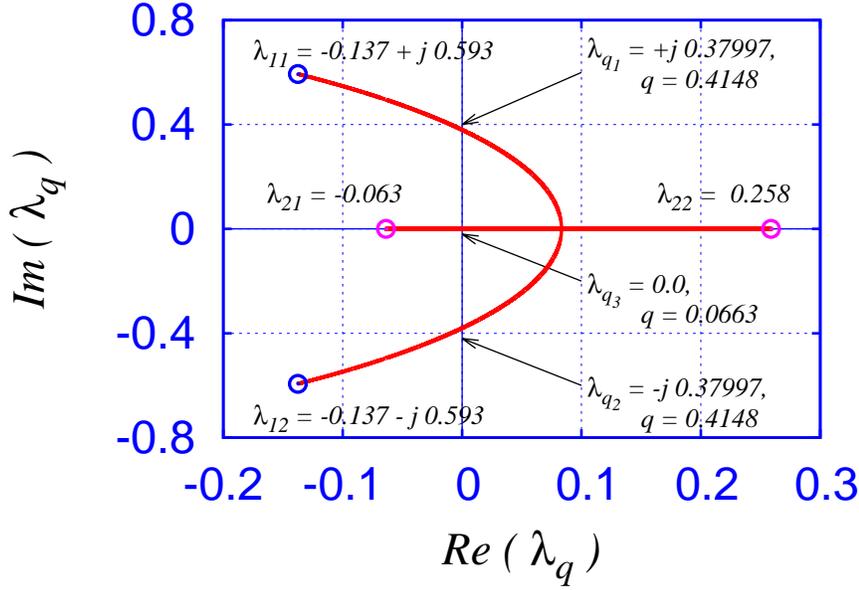,width=5in} 
	\end{center}
	\caption[eigenvalues of the generalised Jacobian] {The one-dimensional
path of the generalised eigenvalues $\lambda_q$'s in the complex plane parametrised by the auxiliary variable $q$, showing the jump of the eigenvalues (indicated by arrows) as a conjugated pair at $q = 0.4148$, signalling a {\textit{discontinuous Hopf}} bifurcation. Also a third crossing of the eigenvalue with the imaginary axis occurring for $q = 0.0663$, indicating a {\textit{multiple crossing}} bifurcation. For the exterme values of the auxiliary variable, namely  $q = 0$ and $q = 1$, the generalised eigenvalues get reduced to the eigenvalues of either of the Jacobian matrices $N_{1,3}$ or $N_2$ in the outer and inner subspaces of the phase space, respectively, and are shown encircled.}
	\label{fig:mmlc_eigen}
\end{figure}

The evolution of non-smooth dynamics near the switching discontinuities may be investigated by setting up a generalised Jacobian matrix $N_q$ at the equilibrium point $E_{0}$, refer \citet{leine06}. This generalized Jacobian matrix is a set-valued matrix and is given as $N_q = qN_2+(1-q)N_{1,3},\;q\in[0,1]$. When written explicitly we have
\begin{equation}
N_q =   \left ( \begin{array}{ccc}
				0  \; \;\;  & \;\;\;1\;\;\;	&  \;0  		 	\\
				0  \; \;\;   &((a_1-a_2)q-a_1)\;    	&  \;1 		 	\\
				0  \; \;\;   &-\beta  		\;	&  -\beta 	\\				 
				\end{array}
		\right).
	\label{eqn:mmlc_genJac}
\end{equation}
 When the auxiliary variable $q = 0$, then from Eq. (\ref{eqn:mmlc_genJac}), we find that $N_q = N_{1,3}$. The eigenvalues $\lambda_q$ in this case therefore reduce to the eigenvalues of the Jacobian matrices $N_{1,3}$ in the outer regions, namely $\lambda_{(\mathbf{1,3})1} = 0.0$ and $\lambda_{(\mathbf{1,3})2,3} = -0.1375 \;\pm\; j\; 0.5935861 $. On the other hand, if $q = 1.0$, then from Eq. (\ref{eqn:mmlc_genJac}), we find that $N_q = N_2$. Hence the eigenvalues $\lambda_q$ reduce to the eigenvalues of the Jacobian matrix $N_2$ in the central region, namely $\lambda_{(\mathbf{2})1} = 0.0$ and $\lambda_{(\mathbf{2})2} = 0.258765$ and $\lambda_{(\mathbf{2})3} = -0.0637645$. These are shown enclosed by circles in Fig. \ref{fig:mmlc_eigen}. 
 
For other intermediate values of the auxiliary variable $q$, the eigenvalues $\lambda_q$ of the generalised Jacobian are set-valued $\lambda_q ,q \in [0,1]$ and form a one-dimensional path in the complex plane. Following this one dimensional path, we find that the eigenvalues become purely imaginary, namely $\lambda_{qi} \simeq \, \pm \,j \, 0.37997$, where $i=1,2$ and $j = \sqrt{-1}$ is the imaginary unit. These generalised eigenvalues $\lambda_{qi}$ are shown by arrows in Fig. \ref{fig:mmlc_eigen} and are found to jump through the imaginary axis as a conjugate pair at $q = 0.41480$, thereby suggesting the occurrence of a discontinuous Hopf bifurcation. This bifurcation, though discontinuous, behaves very much like a continuous Hopf bifurcation found in smooth dynamical systems.

\begin{figure}[htbp]
	\begin{center}
			\psfig{file=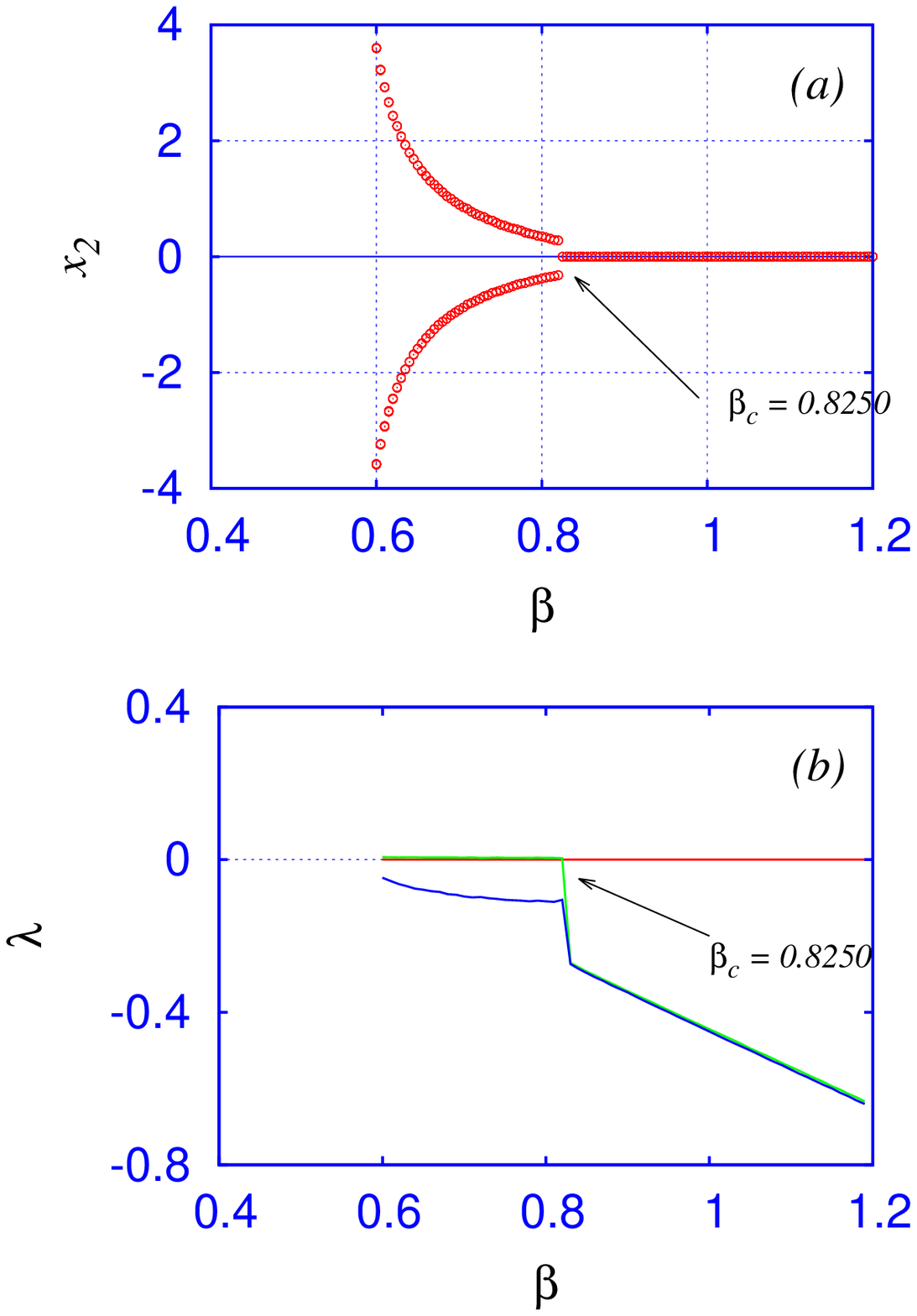,width=5in} 
	\end{center}
	\caption[One Parameter Bifurcation and Lyapunov Spectrum] {(a) The one
parameter bifurcation diagram in the ($\beta-x_2$) plane showing the occurrence of discontinuous Hopf bifurcation causing the fixed points $E_{\pm}$ to loose their stability at the critical value $\beta _c = 0.825$ and resulting in the birth of periodic orbits. (b) The Lyapunov spectrum in the ($\lambda - \beta $) plane corroborating the above inference.  The normalised parameters for the circuit are chosen as $a_{1,3} = -0.55$ and $a_2 = -1.02$.}			
	\label{fig:mmlc_bifspec}
\end{figure}

For the parameters of the memristive MLC circuit mentioned above, the one parameter bifurcation diagram in the ($\beta-x_2$) plane and the Lyapunov spectrum are plotted as a function of $\beta$, for the range $0.6 < \beta < 1.2 $. These are shown in Figs. \ref{fig:mmlc_bifspec} (a) and \ref{fig:mmlc_bifspec} (b) respectively. The bifurcation diagram was obtained by constructing the {\textit{two-sided Poincar\`{e} Map $P_{\pm}$}}, refer \citet{park86}. As the parameter $\beta$ is reduced from $\beta = 1.2$, the bifurcation diagram shows a fixed point behaviour for the system. At a critical value $\beta_c = 0.8250$, the fixed points loose their stability and a bifurcation is observed suggesting the creation of periodic motion. This behaviour continues till $\beta = 0.6$. The Lyapunov spectrum shows that for the control parameter less than the critical value $\beta_c = 0.8250$, the Lyapunov exponents are $\lambda_1 \simeq \lambda_2 = 0.0$ and $\lambda_3 < 0.0$, thereby suggesting that what is observed is in fact quasi-periodic motion.

\begin{figure}[h]
	\begin{center}
			\psfig{file=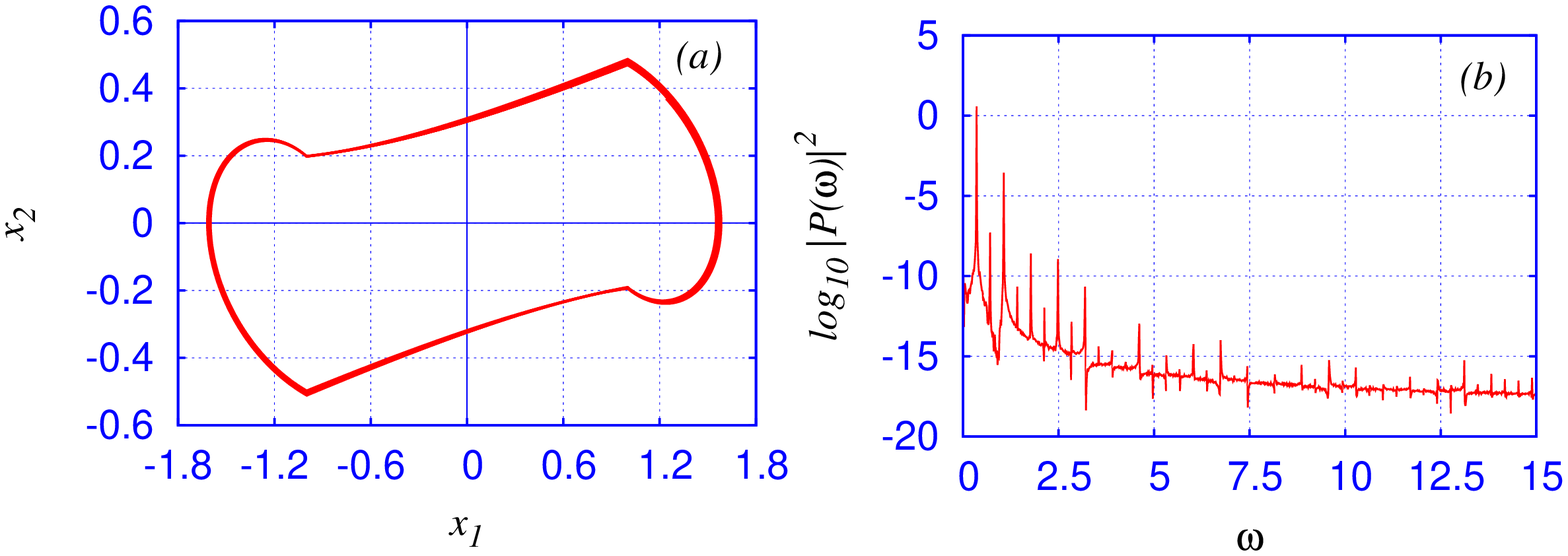,width=6.5in} 
	\end{center}
	\caption[Phase Portrait and the Power Spectrum of periodic orbits]{ (a) The phase portrait showing the quasi-periodic orbits arising due to discontinuous Hopf bifurcation, for parameter values below the critical limit $\beta_c = 0.825$. (b) The corresponding power spectrum of the $x_2$ variable contains Dirac delta peaks, confirming the quasi-periodic behaviour.}
	\label{fig:mmlc_spec}
\end{figure}

Referring back to the phase portraits shown in Fig. \ref{fig:mmlc_limitcycle}, for $\beta_c = 0.8250$, we find that the system has equilibrium points $E_{\pm}$, which are {\textit{stable foci}} in the subspaces $S_1$ and $S_3$ as suggested by the eigenvalues for the Jacobian matrices $N_{1,3}$. As the eigenvalues for the Jacobian matrix $N_2$ suggest an unstable equilibrium, namely a {\textit{saddle}}, the equilibrium point $E_0$ in the subspace $S_2$ repels the trajectories away from it, causing them to cross the switching boundaries $\Sigma_{1,2}$ and $\Sigma_{2,3}$ and settle down to either $E_{+}$ in $S_3$ or to $E_{-}$ in $S_1$ respectively. This is the picture observed for control parameter $\beta$ values less than the critical value $\beta_c = 0.8250$.
\begin{figure}[h]
	\begin{center}
	\psfig{file=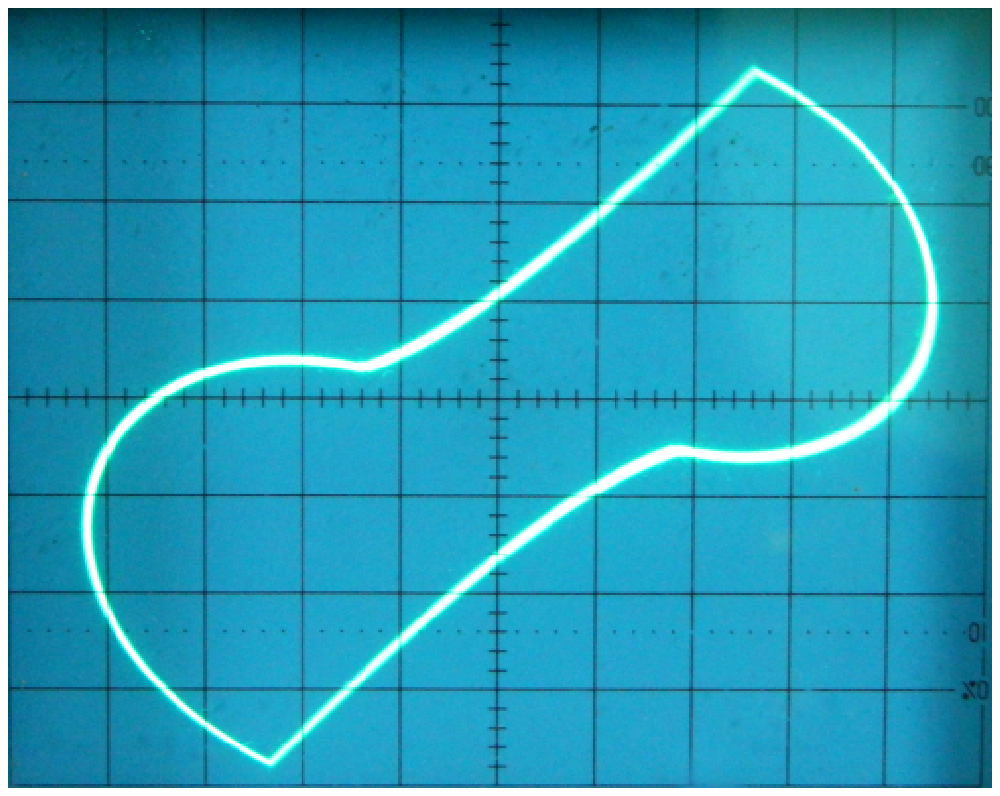,width=2.5in} 
	\psfig{file=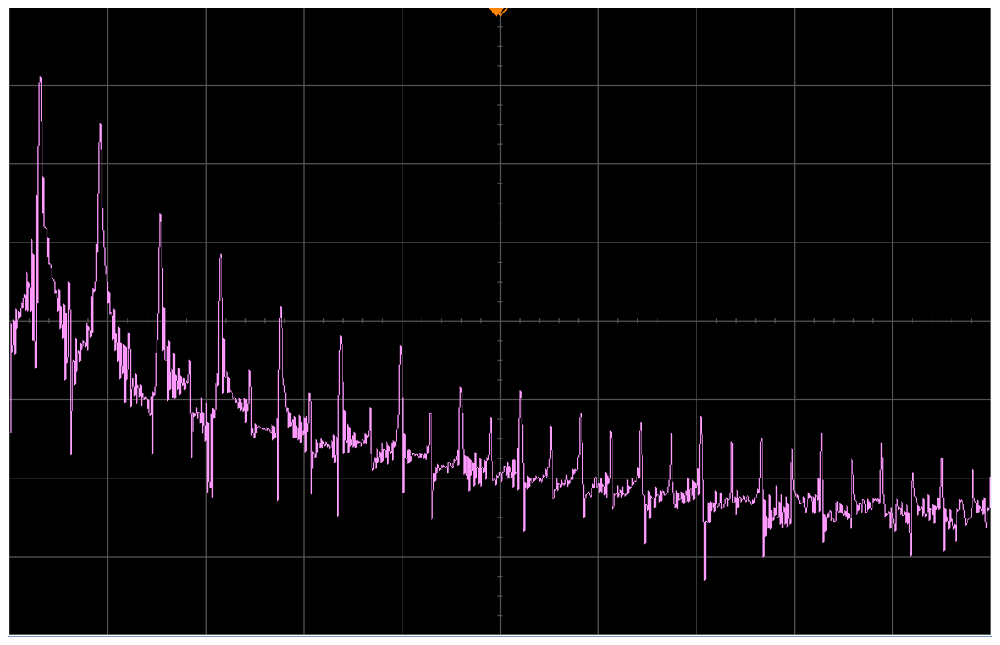,width=3.1in} 
	\end{center}
	\caption[Experimental Phase Portrait showing Hopf Bifurcation and its power spectrum]{Figure showing (a) the experimental phase portrait of the quasi-periodic motion arising due to discontinuous Hopf bifurcation, corresponds well to the  to the numerical phase portrait in Fig. \ref{fig:mmlc_limitcycle}. The experimental values of the circuit elements are L = 300 mH, C = 40 nF and F = 0, that is no external periodic forcing is applied and (b) the power spectrum of the voltage across the capacitor $v_c$, corresponding to the numerical spectrum shown in Fig. \ref{fig:mmlc_spec}, contain Dirac delta peaks, confirming the occurrence of quasi-periodic motion.} 
	\label{fig:expt_mmlc_lc_spec}
\end{figure}
For the control parameter values less than this critical value, $\beta_c = 0.8250$, we observe quasi-periodic behaviour as shown in Fig. \ref{fig:mmlc_spec}(a). These quasi-periodic trajectories pass through all the three subspaces of the state space and enclose the equilibrium points $E_{\pm}$, shown in Fig. \ref{fig:mmlc_limitcycle}, as well. The corresponding power spectrum of the $x_2$ variable, shown in Fig. \ref{fig:mmlc_spec}(b), contains Dirac delta peaks which clearly points to quasi-periodic behaviour. 

\subsection{Explanation of Quasi-periodic Behaviour}
Though the eigenvalues for the generalised Jacobian matrix $N_q$ point to a limit cycle motion alone, the quasi-periodicity exhibited by the system can be explained on the basis of the circuit theoretic properties of the memristor demonstrated in Sec. 2. The discontinuous Hopf bifurcation leads to the birth of periodic motion resulting in self oscillations in the circuit. However the {\textit{linear-time varying}} property of the memristor gives rise to memristor switching. The interaction of the self oscillations and the memristor switching leads to {\textit{memristor induced modulation}} which introduces additional frequencies and their harmonics, thereby causing quasi-periodic motion.

Referring to Fig. \ref{fig:mmlc_eigen} once again, we find that in addition to the the eigenvalues jumping the imaginary axis as a conjugate pair, we find that a third generalised eigenvalue of the system $\lambda_{q3}$ crosses the imaginary axis at the origin for $q =0.0663$. Hence we find a {\textit{pitch-fork}} like bifurcation scenario. This resulting discontinuity induced bifurcation is called {\textit{multiple crossing}} bifurcation \cite{leine06} in the literature, because the eigenvalues cross the imaginary axis more than once during their jump. 

\subsection{Experimental Observations}
The analog circuit of the memrsitive MLC oscillator shown in Fig. \ref{fig:prototype} has been implemented in the laboratory using off the shelf components. Here the integrator, the window comparator as well as the negative impedance converter for the memristive part have been implemented using TL081 operational amplifiers. The electronic switch $AD7510DJ$ has been used for realizing the switching action. The reason for using TL081 is that it operates at high frequency ranges, has high slew rate and does not show hysteretic behaviour. For the integrator part, the parameters are chosen as $ R_1 = 10 K\Omega$, $R_2 = 100 K\Omega$, $R_3 = 100 K\Omega$ and $C_3 = 2.2 nF$. For the window comparator part the output resistance is selected as $R_4 = 10 K\Omega$, while the reference voltages for are fixed as $\pm 1V$. Further we have selected the linear resistances $R_5 = 1450 \Omega$ and $R_6 = 1050 \Omega$, $R_7 = 2 K\Omega$ and $R_8 = 2 K\Omega$ for the negative conductance. The experimental observations were made by employing a Hewlett-Packard Arbitrary Function Generator (33120A) of frequency $15$ MHz, an Agilent Mixed Storage Oscilloscope (MSO6014A) of frequency $100$ MHz and sampling rate of $2$ Giga Samples / seconds. The experimentally obtained phase portrait for quasi-periodic oscillations arising due to Hopf Bifurcation and the experimental power spectrum corresponding to the voltage across the capacitor, $v_c$ are shown in Fig. \ref{fig:expt_mmlc_lc_spec}. The parametric values chosen are $L = 300 mH$, $C = 40 nF$ and when the external force is switched OFF, that is $F = 0$. Obviously we find that the quasi-periodic behaviour results due to discontinuity induced Hopf bifurcations. We find that the experimental phase portrait and the power spectrum agree well with the numerical behaviour shown in Figs. \ref{fig:mmlc_spec}.

\begin{figure}[h]
	\begin{center}
		\psfig{file=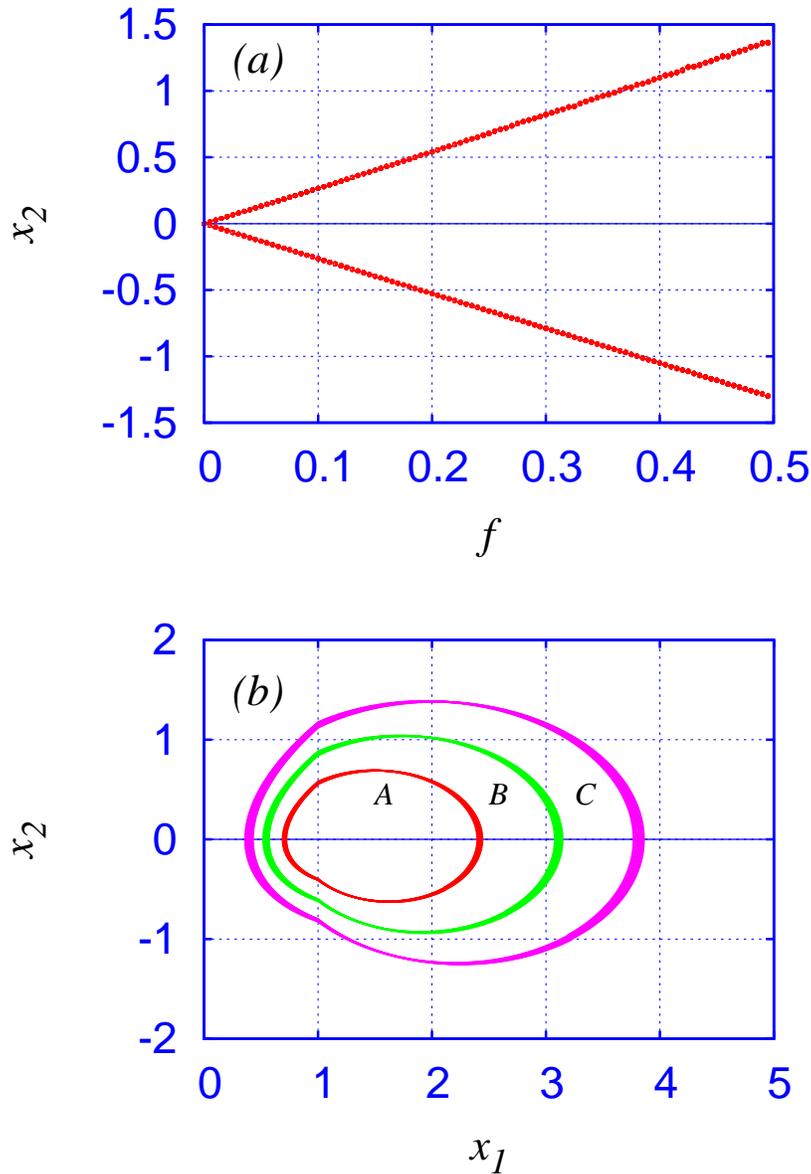,width=5in} 
	\end{center}
	\caption[Amplitude scaling of forced oscillations]{(a) Bifurcation in the ($f-x_2$) plane and (b) Phase portraits in the ($x_1-x_2$) plane showing the amplitude scaling of the orbits due to forced oscillations with the amplitude of the external force. For the forcing amplitudes $f = 0.2$ $0.3$ and $0.4$, the forced periodic orbits are named as A, B and C respectively. We find that the area of these orbits scale as  $A < B < C$ in accordance with the forcing amplitudes.}
	\label{fig:mmlc_force}
\end{figure}

\section{Amplitude Scaling of Forced Oscillations}
In the previous sections we have considered the unforced case of the memrestive MLC circuit, that is the case when $f = 0$. Here, in this section,  we show that when an external sinusoidal forcing is included, the circuit admits forced oscillations that may be quasi-periodic, chaotic or hyperchaotic depending on the vector fields $F_i$'s. 

We have seen that the admissible equilibrium points $E_{\pm}$ for the memristive MLC circuit are asymptotically stable for control parameter values above the Hopf bifurcation value $\beta_c$ in the absence of external force. 
However when external periodic excitation of amplitude $f$ is applied, we find that though the circuit does not undergo any boundary equilibrium bifurcations (BEBs), these admissible equilibrium points $E_{\pm}$ become unstable and give birth to periodic/quasiperiodic oscillations. The amplitudes of these orbits are found to scale linearly with the amplitude of the external force $f$. 

The parameters of the circuit are chosen to be the same as in the autonomous case, that is $a_1 = -0.55$, $a_2 = -1.02$, $\beta = 1.0$ with the frequency of the periodic oscillations as $\omega=0.75$. A bifurcation diagram is drawn by taking the intersections of the periodic orbits with a \textit{two-sided Poincar\`{e} section $P_{\pm}$}. Figure \ref{fig:mmlc_force}(a) shows the bifurcation diagram in the ($f-x_2$) plane wherein the amplitude of the quasi-periodic orbits scale with the amplitude $f$ of the external force for the range $0 < f < 0.5$.  This is shown more explicitly in the phase diagrams in Fig.  \ref{fig:mmlc_force}(b). Here the phase portraits in the ($x_1-x_2$) plane for the forcing amplitudes $f = 0.2$, $0.3$ and $0.4$ are plotted and are named as A, B and C respectively. We find that the area of the orbits scale as $A < B < C$ in accordance with the forcing amplitudes.

\section{Discontinuity Induced Neimark-Sacker Bifurcations}

The memristive MLC circuit which is found to exhibit forced oscillations with a frequency equal to the external forcing frequency $\omega_{ext}$, undergoes  what is called as the {\textit{discontinuity induced secondary Hopf}}  bifurcations or {\textit{discontinuity induced Neimark-Sacker}} bifurcations as the control parameter $\beta$ is varied. Generally  Neimark-Sacker bifurcations are found to give rise to {\textit{two-frequency}} quasi-periodic solutions. However in this circuit, due to the presence of memristor modulation, {\textit{$n$-frequency}} quasi-periodic solutions are generated. According to Floquet theory, the occurrence of this type of bifurcations can be explained by setting up the {\textit{variational equation}} for the system and determining the {\textit{characteristic multipliers}} of the {\textit{monodromy matrix}}.

\begin{figure}[htbp]
	\begin{center}
		\psfig{file=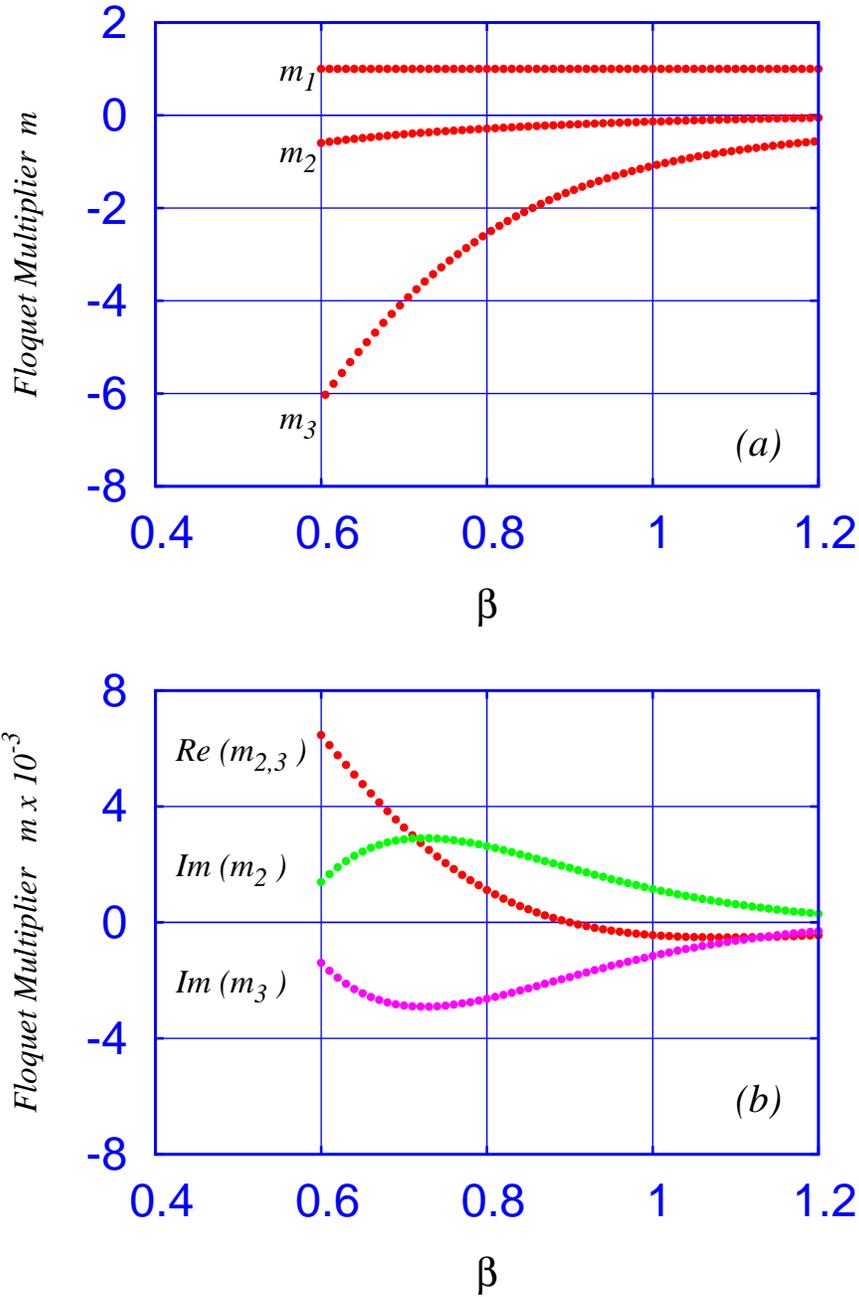,width=5.5in} 
	\end{center}
	\caption[Variation of the Floquet multipliers with the control parameter]{Variation of (a) the characteristic multipliers $m_i$'s of the 	monodromy matrix for the range $0.6 < \beta < 1.2$ and of (b) the generalised characteristic multipliers $m_{q_i}$'s for the same range. 	In both the cases one of the multipliers is unity. The crossing of the real part of $m_{q_{2,3}}$ at the critical value $\beta_m = 0.9$ signals discontinuity induced Neimark-Sacker bifurcation.}	
	\label{fig:mmlc_neimarck}
\end{figure}

The variational equation is a matrix-valued time-varying linear differential equation. It is the linearisation of the vector field along the trajectory of the system. For smooth dynamical systems defined by
\begin{equation}
\dot{\mathbf{x}} = F(\mathbf{x},t), \qquad \mathbf{x}(t_0) = \mathbf{x}_0
	\label{eqn:smooth_sys}
\end{equation}
the variational equation is given as, refer \citet{park86},
\begin{eqnarray}
\dot{\Phi}_t(\mathbf{x_0},t_0) & = & D_xF(\phi_t(\mathbf{x_0},t_0),t)\Phi_t(\mathbf{x_0},t_0), \nonumber \\ 
\Phi_{t_0}(\mathbf{x_0},t_0) & = & I. 
	\label{eqn:variational}
\end{eqnarray}
where $\phi_t(\mathbf{x_0},t_0)$ is the {\textit{flow function}}. Here $\phi_t(\mathbf{x_0},t_0)$ is the solution of the Eq. (\ref{eqn:smooth_sys}), that is  $\phi_t(\mathbf{x_0},t_0) = x_0$. Further
\begin{eqnarray}
\Phi_t(\mathbf{x_0},t_0):=D_{x_0}\phi_t(\mathbf{x_0},t_0),\nonumber \\ 
\Phi_{t_0}(\mathbf{x_0},t_0):=D_{x_0}\phi_{t_0}(\mathbf{x_0},t_0). 
\end{eqnarray}
Here $D_x$ stands for vector differentiation and $I$ is the identity matrix. 

The eigenvalues  of the monodromy matrix are called {\textit{characteristic multipliers}} or {\textit{Floquet multipliers}} and are usually denoted as $m_i$'s. Just as the eigenvalues $\lambda_i$'s of the Jacobian matrix determine the stability of the equilibrium points of the system, the characteristic multipliers $m_i$'s of the monodromy matrix determine the stability of the periodic solutions.

For a piecewise-smooth system, having a single switching manifold $\Sigma_{1,2}$, defined by the zero vlaue of the scalar function $H(\mathbf{x})$, the state equations are
\begin{equation}
\dot{\mathbf{x}} = 
		\left\{
			\begin{array}{ll}
				F_1(\mathbf{x},t) \qquad \textrm{if}\,\,H(\mathbf{x}) < 0, \\
				F_2(\mathbf{x},t) \qquad \textrm{if}\,\,H(\mathbf{x}) > 0.
			\end{array}
		\right.
		\label{eqn:ns_deg1}
\end{equation}
As the vector fields $F_i$'s for $i = 1,2$ are discontinuous at the switching boundaries  $\Sigma_{1,2}$, the gradient $ D_xF(\phi_t(\mathbf{x_0},t_0),t)$ does not exist when $x(t) \in \Sigma_{1,2}$. The crossing of the switching boundary $\Sigma_{1,2}$ by the solution $x(t)$ at an instance of time, say $t_p$ causes the monodromy matrix to jump at this switching boundary. 

Using the concept of generalised differential of Clarke, we saw that the generalisation of a Jacobian matrix $N_{1,2}$ around an equilibrium point of a non-smooth system gave rise to a set-valued generalised Jacobian matrix $N_q$, $q \in [0,1]$, whose eigenvalues are also set-valued, that is $\lambda_q$s, $q \in [0,1]$. These set-valued eigenvalues $\lambda_q$s were able to explain correctly the behaviour of the non-smooth system at the switching boundaries. 

The monodromy matrix is a linearisation of the periodic solution and therefore for all practical purposes be considered as a Jacobian matrix for periodic orbits. Hence in a manner similar to the construction of a generalised Jacobian, \citet{leine06} had extended the concept of generalisation to the monodromy matrix also, constructing thereby a set-valued monodromy matrix, say $Q$. This set-valued monodromy matrix is called in the literature of nonsmooth systems as the {\textit{Saltation Matrix}}. The eigenvalues of the Saltation matrix will also be set-valued and are called the generalised characteristic multipliers $m_{q_i}$'s. The crossing of the imaginary axis of these characteristic multipliers as a conjugate pair in the eigen space will then cause the periodic orbits to loose their stability and give birth to further closed invariant curves with frequencies different from that of the parent periodic orbit in the state space. Depending on the relation of the frequencies of the  new orbits with that of the parent orbit, the newly formed orbits may be periodic or quasi-periodic.

The memristive MLC circuit being a piecewise-smooth system with two discontinuity boundaries $\Sigma_{1,2}$ and $\Sigma_{2,3}$ defined by the zeros of two scalar functions $H_1(\mathbf{x},\mu)$ and $H_2(\mathbf{x},\mu)$, its state equations, refer Eq. (\ref{eqn:smooth_odes}), are given as a set of smooth ODEs,
\begin{equation}
\dot{\mathbf{x}}(t) = 
	\left\lbrace	\begin{array}{lcccl}
		F_2(\mathbf{x},\mu), & H_1(\mathbf{x},\mu) \geq  0 & \& & H_2(\mathbf{x},\mu) <0,& x \in S_2,  \\
					&					  &     &                &      \\
		F_{1,3}(\mathbf{x},\mu), & H_1(\mathbf{x},\mu)<0 &\& & H_2(\mathbf{x},\mu)\geq 0,& x\in S_{1,3},
		\end{array}
	\right.
	\label{eqn:nssmooth_odes}
\end{equation}
where $\mu$ denotes the parameter dependence of the vector fields and the scalar functions. The vector fields $F_i$'s are
\begin{equation}
 F_i(\mathbf{x},\mu) =  \left (	\begin{array}{c}
				x_2		\\
				-a_i x_2	+x_3					\\
				-\beta x_2 -\beta x_3 +f sin(\omega x_4)	\\	
					1 
				\end{array}
		\right ), \text{i\;=\;1,2,3}.
\label{eqn:nsvect_field}
\end{equation}
The variational equation for the system was set up by substituting Eq. (\ref{eqn:nssmooth_odes}) in Eq. (\ref{eqn:variational}), where $D_xF(\phi_t(\mathbf{x_0},t_0),t)$ is given  as 
\begin{equation}
D_xF(\phi_t(\mathbf{x_0},t_0),t) =   \left ( \begin{array}{ccc}
				0      & 1    	&  0  		 \\
				0      &W    &  1 		 \\
				0      &-\beta  &  -\beta 	\\				 
				\end{array}
		\right), \nonumber
	\label{eqn:F_vecdiff}
\end{equation}
with $W = q(a_1-a_2)-a_1$, $q \in [0,1]$. 
The solution of this variational equation was obtained numerically. As this involves both $\phi_t(\mathbf{x_0},t_0)$ and $\Phi_t(\mathbf{x_0},t_0$, the variational equation was appended to the original system to obtain the combined system 
\begin{equation}
\left \{
	\begin{array}{c}
	\dot{\mathbf{x}} \\
	\dot{\Phi}
	\end{array}
\right \} = 
\left \{
	\begin{array}{c}
		F_i(\mathbf{x},t) \\
		D_xF_i(\mathbf{x},t)\Phi
	\end{array}
\right \}.
	\label{eqn:combined_varitional}	
\end{equation}
This combined system of equations was solved simultaneously by numerical integration for a time $T = \left ( \dfrac{2 \pi}{\omega} \right ) $  assuming the initial conditions
\begin{equation}
\left \{
	\begin{array}{c}
	\mathbf{x}(t_0) \\
	\Phi(t_0)
	\end{array}
\right \} = 
\left \{
	\begin{array}{c}
		\mathbf{x_0} \\
		I
	\end{array}
\right \}.
	\label{eqn:combined_initial}	
\end{equation}
where $I$ is an identity matrix. 

\begin{figure}[htbp]
	\begin{center}
			\psfig{file=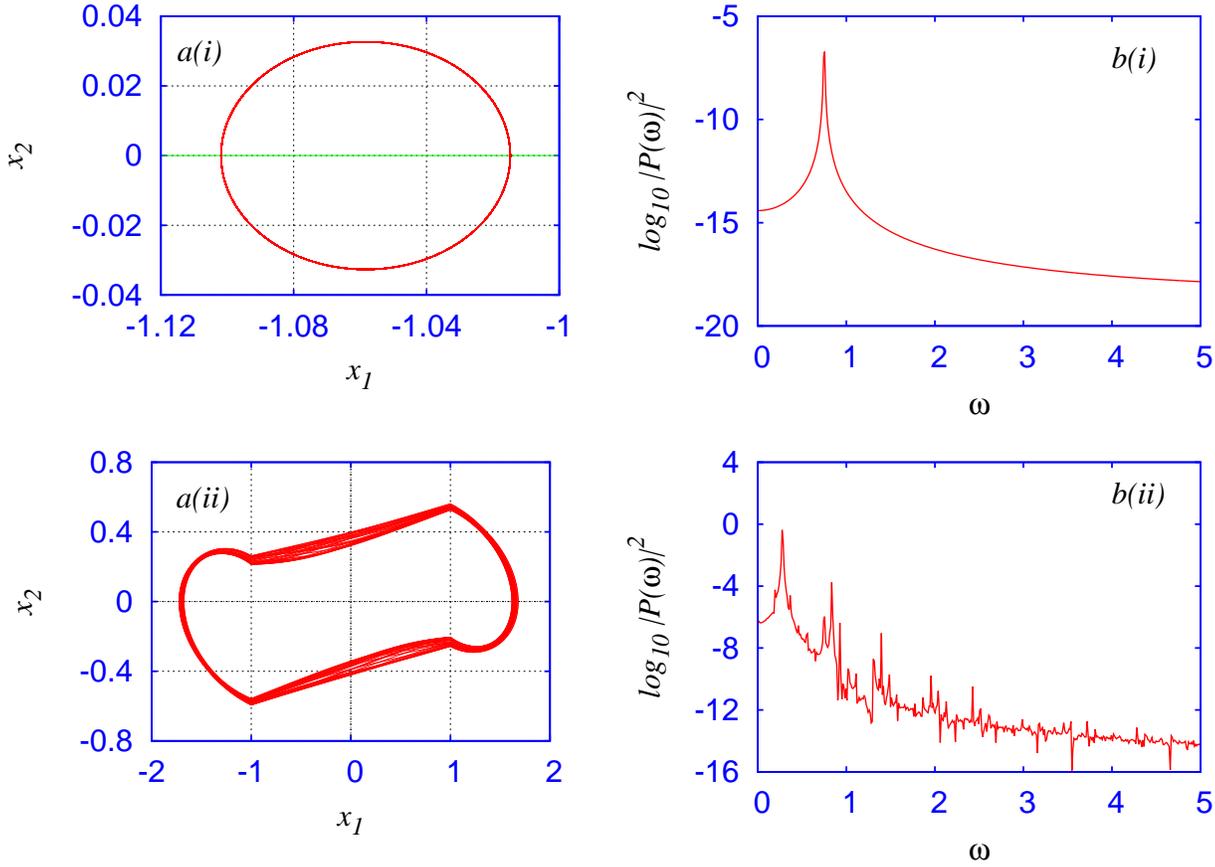,width=7in} 
		\end{center}
	\caption[Phase portraits and power spectra before and after a Neimark -Sacker bifurcation]{Phase portraits a(i) and a(ii) and their corresponding power spectra b(i) and b(ii) for the memristive MLC circuit before and after a Neimark-Sacker bifurcation. The single peak at a frequency $\omega = 0.755310$ very close to the frequency of the external force and a peak power of $log_{10}|P(\omega)|^2 = -6.714105$ of the power spectrum in b(i) represents that of the forced oscillations while the Dirac delta peaks of the power spectrum in b(ii) confirms the quasi-periodic behaviour.} 
	\label{fig:phase_neimarck}
\end{figure}

The normalised circuit parameters are assumed to be the same as before with $a_{1,3} = -0.55$, $a_2 = -1.02$, with the normalised angular frequency of the external periodic forcing as  $\omega_{ext}=0.75$ and the amplitude of the external force as $f = 0.01$. The parameter $\beta$ is assumed as the control parameter for the system and the auxiliary variable is set as $q = 0.5$. The eigenvalues of this generalised monodromy matrix so obtained, namely $m_{q_i}$'s, are also set valued $q \in [0,1]$. 

\begin{figure}
		\begin{center}
		\psfig{file=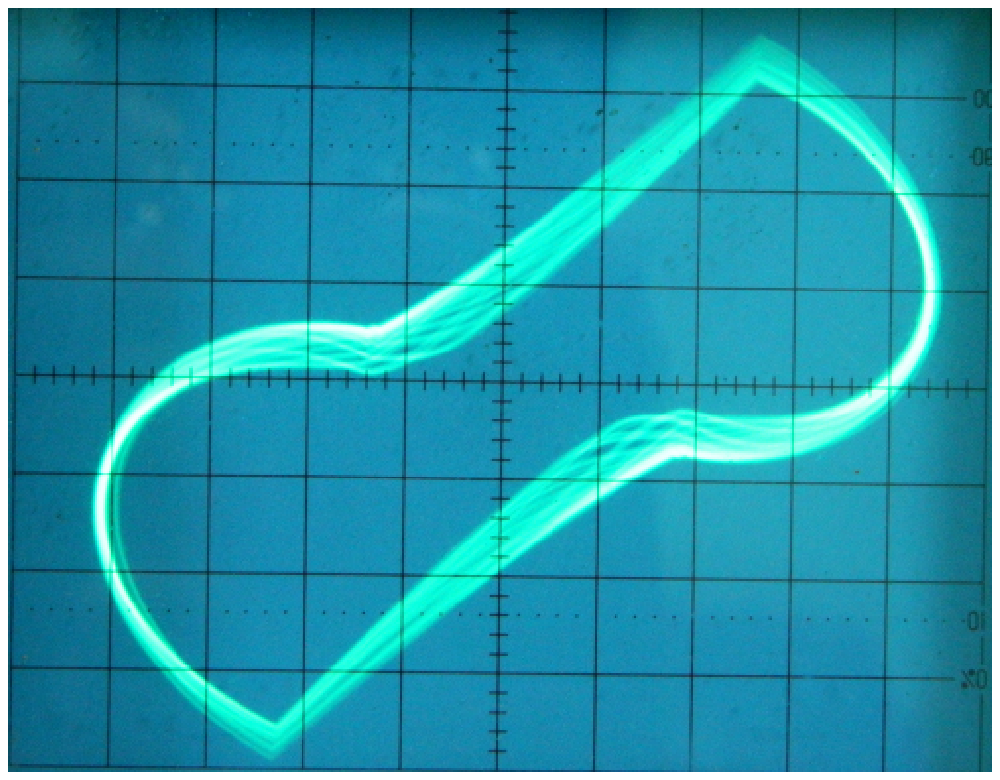,width=3in} 
		\psfig{file=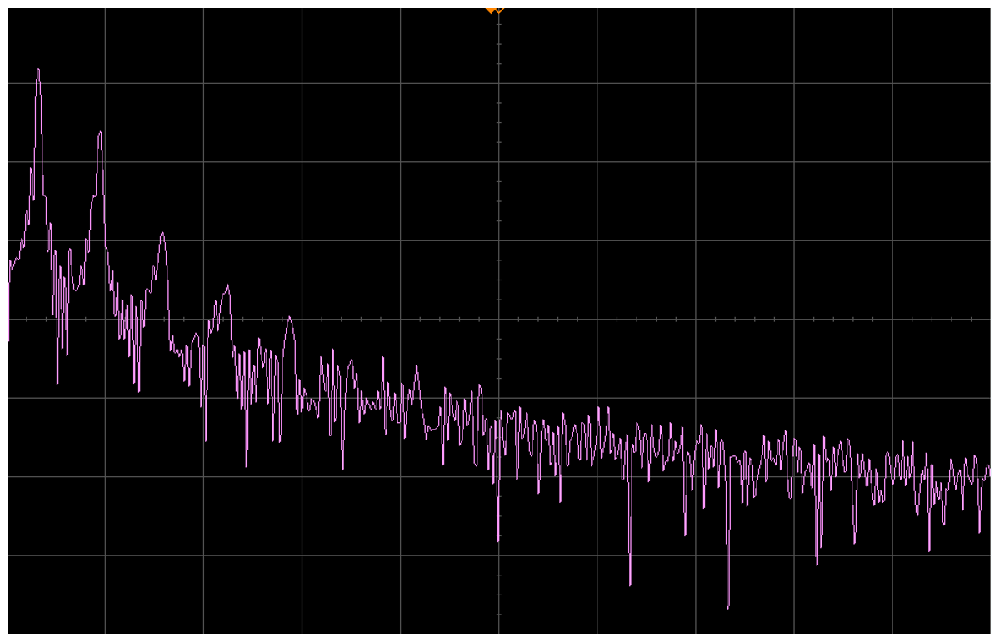,width=3.67in} 
		\end{center}
		\caption[Experimental Phase Portrait showing Neimark-Sacker Bifurcation and its power spectrum]{Figure showing (a) the experimental phase portrait in the $(\phi-v_c)$ plane of the quasi-periodic motion arising due to discontinuous Neimark-Sacker bifurcation, corresponding to the numerical phase portrait in Fig. \ref{fig:phase_neimarck} a(ii). The experimental values of the circuit elements are $L = 300 mH$, $C = 40 nF$, amplitude $F = 380 mV_{pp}$ (peak-to-peak voltage) and frequency $f = 1.972KHz$ for the externally applied force and (b) the power spectrum of the voltage across the capacitor $v_c$, corresponding to the numerical spectrum shown in Fig. \ref{fig:phase_neimarck} b(ii), depicting Dirac delta peaks, confirming the occurrence of quasi-periodic motion.} 
	\label{fig:expt_mmlc_nsbif_spec}
\end{figure}

The variations of the Floquet multipliers as a function of the control parameter in the range $0.6 < \beta < 1.2$ for two different cases were investigated. Firstly the monodromy matrix was set up in a manner similar to that of a smooth system. Secondly the generalised monodromy matrix was constructed by applying the concept of Clarke's generalised differential. These variations are shown in Fig. \ref{fig:mmlc_neimarck} (a) and (b) respectively. As is expected from Floquet theory, one of the characteristic multipliers is always unity, that is $m_1 = m_{q_1} = 1$.  In Fig. \ref{fig:mmlc_neimarck} (a) the other two characteristic multipliers are negative $m_{2,3} = <0$. This tells that the periodic orbit arising due to forced oscillations is highly stable and hence no further bifurcations or any new attractors are possible for the entire range $0.6 < \beta < 1.2$ of the control parameter. Hence the usual Floquet theory does not predict any Neimark-Sacker Bifurcation in the memristive MLC circuit.

However in Fig. \ref{fig:mmlc_neimarck} (b) the set-valued characteristic multipliers $m_{q_{2,3}}$ cross the imaginary axis as a complex conjugate pair at a critical value of the control parameter $\beta_m = 0.9$. Above this critical value of $\beta$ the real part of $m_{q{2,3}}$ is negative while it crosses the zero axis at the critical value and becomes positive thereafter. This shows that for control parameter above the critical value, that is for $\beta > \beta_m$ the periodic orbit is stable, while below it, the periodic orbit losses its stability giving birth to a newer periodic orbit as a result of {\textit{discontinuity induced Neimark-Sacker}} bifurcation. The frequency of the new orbit may be different from that of the parent one. Further the memristor modulation causes once again the periodic orbit to become a $n -$ frequency quasi-periodic attractor.

The stable periodic orbit due to forced oscillations is shown in the phase portrait in Fig. \ref{fig:phase_neimarck} a(i). Its corresponding power spectrum shown in Fig \ref{fig:phase_neimarck} b(i) contains just a single peak at a frequency $\omega = 0.755310$ very close to the frequency of the external force ($\omega_{ext} = 0.75$) and a peak power of $log_{10}|P(\omega)|^2 = -6.714105$. Similarly the phase portrait in Fig. \ref{fig:phase_neimarck} a(ii) and the Dirac delta peaks in its corresponding power spectrum in Fig. \ref{fig:phase_neimarck} b(ii) show clearly the quasi-periodic nature of the attractor.

As in the case of Hopf Bifurcation, the experimental observation of the Neimark-Sacker Bifurcation was also obtained. For this, the parameters of the memristive part of the circuit as well as the inductance and capacitance values, namely $L = 300 mH$ and $C = 40 nF$ were retained as were fixed earlier for observing Hopf Bifrucation. However when the amplitude and frequency of the external force were increased, we observed Neimark-Sacker bifurcation taking place. The experimental phase portrait in the $(\phi-v_c)$ plane of the quasi-periodic motion when the amplitude of the external force $F = 380 mV Vpp$ with its frequency being $f = 1.972 KHz$ and the power spectrum of the voltage $v_c$ across the capacitor are shown in Fig. \ref{fig:expt_mmlc_nsbif_spec}. These correspond well to the phase portrait in the $(x_1-x_2)$ plane shown in Fig. \ref{fig:phase_neimarck} a(ii) and the power spectrum of the $x_2$ normalised variable shown in Fig. \ref{fig:phase_neimarck} b(ii) respectively.

\section{Conclusion}
In this paper we have considered the memristive Murali-Lakshmanan-Chua circuit as a non-smooth dynamical system and studied its peculiar behaviour arising due to the piecewise continuous nature of the memristor, its switching ability, time variation properties and modulation characteristic. We have done this by identifying and setting up switching manifolds, reformulating the system equations as a set of smooth ODEs and by construction of discontinuity mapping corrections such as the Poincar\'{e} Discontinuity Mapping and Zero Time Discontinuity Mapping corrections at these switching boundaries. The details of these discontinuity mapping corrections can be referred from our earlier work, \cite{icha13}. We have found out the equilibrium points admitted by this system, evaluated their stability. The discontinuity induced bifurcations occurring at switching manifolds were analysed. We predicted using Clarke's generalised Jacobian theory the occurrence of discontinuous Hopf bifurcations in the unforced case and using proper modifications of the Floquet theory the existence of Neimark-Sacker bifurcations in the system in the presence of the external forces. We have verified these with the help of numerical simulations as well as real hardware experiments. To the best of our knowledge it is for the first time that discontinuity induced Hopf and Neimark-Sacker bifurcations were identified and studied in a memristive piecewise continuous system.

\section{References}

%\end{multicols}
\nonumsection{Acknowledgments} 
\noindent
This work has been supported by a NASI Platinum Jubilee Senior Scientist fellowship to M.L.\\

\appendix{\label{App:A}}
\nonumsection{Piecewisesmooth Continuous Systems}
A dynamical system may be defined as a deterministic mathematical prescription for evolving the state of a physical entity forward in time. If this time evolution arises as a smooth function of the arguments of the dynamical system, then it is termed as a \emph{smooth dynamical system}. However if the time evolution of the dynamics of the system arises as a nonsmooth function of the arguments, then the dynamical system is termed as a \emph{nonsmooth} or \emph{piecewise smooth} system. The piecewise smooth systems are found to posses one or many discontinuity boundaries. The \emph{discontinuity boundaries} are defined as the sets $\Sigma_{i,j}$ that separate the phase space of a piecewise smooth system into different regions \cite{nord91}. They are also referred to in literature as \emph{discontinuity sets} or  \emph{switching manifolds}. They are identified by the zeros of some scalar functions, say $H(\mathbf{x})$. If the piecewise smooth systems take on a continuous time evolution behaviour at the discontinuity boundaries, then the systems are called as \emph{piecewise smooth continuous systems}. \\

%------------------------
\begin{figure}[h]
	\begin{center}
			\psfig{file=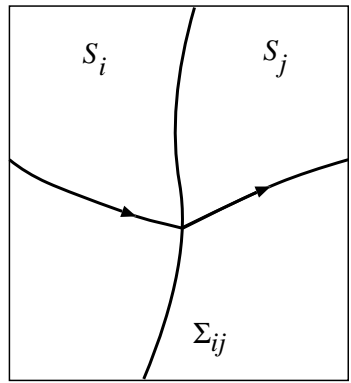} 		 
	\end{center}	
	\caption[PWS Flow] {Schematic illustration of the trajectories of a piecewise-smooth flow crossing the switching manifold $\Sigma_{i,j}$ from the subspace $S_i$ to $S_j$}
	\label{fig:pws_flow1}
\end{figure}
%------------------------
Generally the piecewise smooth continuous systems are described by a finite set of continuous ODEs
\begin{equation}
\dot{\mathbf{x}} = F_i(\mathbf{x},\mu), \qquad \textrm{for} \,\, \mathbf{x} \in S_i
	\label{eqn:pw_ode}
\end{equation}
where $S_i$'s are the subspaces of the phase space created by one or more \emph{discontinuity boundaries} $\Sigma_{i,j}$'s as shown in Fig. \ref{fig:pws_flow1}. Let us consider a piecewise-smooth system having a single discontinuity set $\Sigma_{1,2}$ identified by the zeros of the scalar function $H(\mathbf{x},\mu)$, such that Eq. (\ref{eqn:pw_ode}) can be explicitly written as
\begin{equation}
\dot{\mathbf{x}} = \left \{
						\begin{array}{ll}
							F_1(\mathbf{x},\mu),\; \textrm{if}\; H(\mathbf{x},\mu)<0, \;\mathbf{x} \in S_1   \\
							F_2(\mathbf{x},\mu),\; \textrm{if}\; H(\mathbf{x},\mu)>0, \; \mathbf{x} \in S_2
						\end{array}
					\right.
		\label{eqn:pw_deg}
\end{equation}
This system will have 
\begin{enumerate}
\item
a degree of smoothness \emph{one} if $F_1(\mathbf{x},\mu) \neq F_2(\mathbf{x},\mu)$ at $\mathbf{x} \in \Sigma_{1,2}$.
\item
a degree of smoothness \emph{two} if $F_1(\mathbf{x},\mu) = F_2(\mathbf{x},\mu)$, but $\dfrac{\partial F_1}{\partial x} \neq \dfrac{\partial F_2}{\partial x}$ at $\mathbf{x} \in \Sigma_{1,2}$.
\item
a degree of smoothness \emph{three} if $F_1(\mathbf{x},\mu) = F_2(\mathbf{x},\mu)$ and $\dfrac{\partial F_1}{\partial x} = \dfrac{\partial F_2}{\partial x}$, but $\dfrac{\partial^2 F}{\partial x^2} \neq \dfrac{\partial^2 F}{\partial x^2}$ at $\mathbf{x} \in \Sigma_{1,2}$.
\end{enumerate}
If the degree of smoothness of a piecewise-smooth continuous system is \emph{one}, then the system is called a \emph{Filippov System} \cite{flip64,flip88}. Filippov systems admit a complex type of behaviour called \emph{sliding motion}. However if the degree of smoothness of the system is greater than or equal to \emph{two}, then the system is called as a \emph{Piecewise-smooth Continuous Flow} or simply a \emph{Flow}. A large number of piecewise smooth continuous systems belong to either of these types. For example the familiar Chua's oscillator and the MLC oscillator, both for long considered as paradigmatic examples in the study of chaos, can behave either as a Filippov system or as a piecewise-smooth continuous flow under appropriate conditions. The discontinuity sets in the Chua and MLC oscillator arise because of the piecewise-linear nature of their nonlinear resistive element, namely the Chua's diode. However the chaotic dynamics exhibited by these systems is attributed to a sequence of smooth bifurcations. The presence of discontinuity boundaries in these systems seem to affect only the geometrical structure of the chaotic attractor, see \cite{dib08}.
%--------------------- 
\nonumsubsection{Discontinuity Induced Bifurcation (DIB)}

A discontinuity-induced bifurcation is said to occur at a particular parameter value if there exists an arbitrary small perturbation that causes the system to lose its piecewise-structural equivalence \cite{dib08a}. Different types of discontinuity-induced bifurcations (DIBs) are found to arise, such as border collision in maps, boundary equilibrium bifurcations, grazing bifurcations, sticking and sliding bifurcations and border intersection bifurcations in piecewise smooth continuous systems. These bifurcations are referred to in Russian and East European literature \cite{feigin70,feigin94} as \emph{C-Bifurcations} . The Russian letter '\emph{C}' pronounced as '\emph{S}' stands for \emph{sewing}, as one sews together two different trajectory segments on either side of the discontinuity boundary.
%-------------------------

\nonumsubsection{Equilibrium Points of a Piecewise Smooth Continuous System}
A piecewise-smooth continuous flow will have two types of equilibrium points, namely (i) admissible equilibrium point and (ii) boundary equilibrium point.
A point $x \in S$ is an \emph{admissible equilibrium point} of Eq. (\ref{eqn:pw_ode}) , if
\begin{eqnarray}
F_1(\mathbf{x},\mu) = 0, \\ \nonumber
H(\mathbf{x},\mu) \text{:} = \lambda_1 < 0. 
	\label{eqn:pw_admissible1}
\end{eqnarray}
or
\begin{eqnarray}
F_2(\mathbf{x},\mu) = 0,  \\ \nonumber
H(\mathbf{x},\mu) := \lambda_2 < 0. 
	\label{eqn:pw_admissible2}
\end{eqnarray}
This admissible equilibrium point is also known as a \textit{virtual equilibrium} point.
\\
A point $\hat{\mathbf{x}}$ is a \emph{boundary equilibrium point} of Eq. (\ref{eqn:pw_ode}), if
\begin{eqnarray}
F_1(\hat{\mathbf{x}},\mu) = 0, \qquad \textit{or} \qquad F_2(\hat{\mathbf{x}},\mu) = 0, \nonumber  \\
H(\hat{\mathbf{x}},\mu) = 0.
		\label{eqn:pw_BEB1}
\end{eqnarray}

\nonumsubsection{Boundary Equilibrium Bifurcations (BEB)}
The boundary equilibrium bifurcations are analogous to the border-collision bifurcations in maps. A piecewise-smooth system is said to undergo a boundary equilibrium bifurcation \cite{dib08}, at $\mu = \mu^*$ if there exists a point $x^*$ such that for all values of $i = 1,2,3$:
\begin{eqnarray}
F_i(\mathbf{x^\ast},\mu^\ast) & = &  0 \nonumber \\ 
H_i(\mathbf{x^\ast},\mu^\ast) & = & 0 \nonumber \\
det(F_{i,\mathbf{x}}) & \neq & 0 \nonumber \\
P(\mathbf{x^\ast},\mu^\ast) & \neq  & 0,
	\label{eqn:BEB}
\end{eqnarray}
where
\begin{equation}
P(\mathbf{x^\ast},\mu^\ast) = H_{i,\mu}(\mathbf{x^\ast},\mu^\ast)-H_{i,\mathbf{x}}(\mathbf{x^\ast},\mu^\ast)  \left[F_{i,x}^{-1}F_{i,\mu} \right ](\mathbf{x^\ast},\mu^\ast).
\end{equation}
Here the second subscripts denote the derivatives of the function with respect to the variables represented by those subscripts, namely 
\begin{eqnarray}
H_{i,x}(\mathbf{x}^\ast,\mu ^\ast) & = & \left. \dfrac{\partial H_i(\mathbf{x},\mu)}{\partial x} \right \rvert_{\mathbf{x}=\mathbf{x}^\ast,\mu = \mu ^\ast}  \nonumber \\
H_{i,\mu}(\mathbf{x}^\ast,\mu ^\ast) & = &\left. \dfrac{\partial H_i(\mathbf{x},\mu)}{\partial \mu} \right \rvert _{\mathbf{x}=\mathbf{x}^\ast,\mu = \mu ^\ast} \nonumber \\
F_{i,x}(\mathbf{x}^\ast,\mu ^\ast) & = &\left. \dfrac{\partial F_i(\mathbf{x},\mu)}{\partial x} \right \rvert _{\mathbf{x}=\mathbf{x}^\ast,\mu = \mu ^\ast} \nonumber \\
F_{i,\mu}(\mathbf{x}^\ast,\mu ^\ast) & = &\left. \dfrac{\partial F_i(\mathbf{x},\mu)}{\partial \mu} \right \rvert _{\mathbf{x}=\mathbf{x}^\ast,\mu = \mu ^\ast}.
\end{eqnarray}
The first two equations give the defining conditions for a boundary equilibrium bifurcation to occur at $\mathbf{x^{\ast}}$ for a parameter value $\mu = \mu^*$. The third equation ensures that $\mathbf{x^{\ast}}$ is an \textit{isolated hyperbolic} equilibrium point for the vector fields $F_i$'s. The final equation is a non-degeneracy condition with respect to the parameter $\mu$, that the admissible branches of the equilibria cross the bifurcation point $\mathbf{x^{\ast}}$ at $\mu = \mu^*$. 

Linearising the system described by Eq. (\ref{eqn:pw_deg}) about its boundary equilibrium point, $\mathbf{x}=0$, $\mu = 0$, we have
\begin{eqnarray}
N_1\; \mathbf{x^-} +M_1\;\mu & = & 0, \nonumber \\
C^T \;\mathbf{x^-} +D\;\mu & = & \lambda^-,
	\label{eqn:beba1}
\end{eqnarray}
and
\begin{eqnarray}
N_2\;\mathbf{x^+} +M_2 \;\mu & = & N_1\mathbf{x}^+ M_1\mu + E\lambda^+ = 0, \nonumber \\
C^T\mathbf{x^+} +D\;\mu & = & \lambda^+,
		\label{eqn:beba2}
\end{eqnarray}
where $N_1 = F_{1,x}, M_1 = F_{1,\mu}$, $C^T=H_x$, $D = H_{\mu}$ and $\mathbf{x}^+(\mu)$ and $\mathbf{x}^-(\mu)$ are the branches of the equilibria, before and after the crossing of the discontinuity boundaries at $\mu = \mu^\ast$in the sub-regions $S_1$ and $S_2$ of the vector fields $F_1(\mathbf{x},\mu))$ and $F_2(\mathbf{x},\mu)$ respectively.

From the above relations, the conditions for the occurrence of boundary equilibrium bifurcations (BEBs) can be inferred. 

\begin{description}
\item
{\bfseries{Persistence}}\\
A persistence refers to the change of the nature of the equilibrium point from  being admissible to virtual and vice versa upon crossing of the discontinuity boundary. This bifurcation is observed if the following conditions are satisfied
\begin{eqnarray}
det(N_1) & \neq & 0, \nonumber  \\
D-C^TN_1^{-1}M_1 & \neq & 0, \\
1+C^TN_1^{-1} E &> &0.
		\label{eqn:persistence}
\end{eqnarray}
\item
{\bfseries{Non-smooth Fold}}\\
A non-smooth fold refers to the conversion collision of two branches of \emph{admissible equilibria} upon intersection of the discontinuity boundary to two branches of \emph{virtual equilibria} and vice versa. This bifurcation is observed if 
\begin{eqnarray}
det(N_1) & \neq & 0, \nonumber \\
D-C^TN_1^{-1}M_1 & \neq & 0, \\
1+C^TN_1^{-1} E &< &0.
		\label{eqn:nonsmooth_fold}
\end{eqnarray}
\end{description}
%==============================

\appendix{\label{App:B}}
\nonumsection{Clarke's Theory of generalised differentials and generalized Jacobians}
Let us consider a smooth continuous function $f(x)$. Then the classical derivative of this function at any point $x$ can be defined by the tangent line to the graph of $f(x)$ at that point $x$, namely
\begin{equation}
f'(x) = \dfrac{\partial f(x)}{\partial x} =\; \stackrel{\mathit{limit}}{_{y \rightarrow x}} \dfrac{f(y)-f(x)}{y-x}.
	\label{eqn:classical_deriv}
\end{equation}

\begin{figure}[hbtp]
	\centering
		{\includegraphics[width=0.25\linewidth,angle = 270]{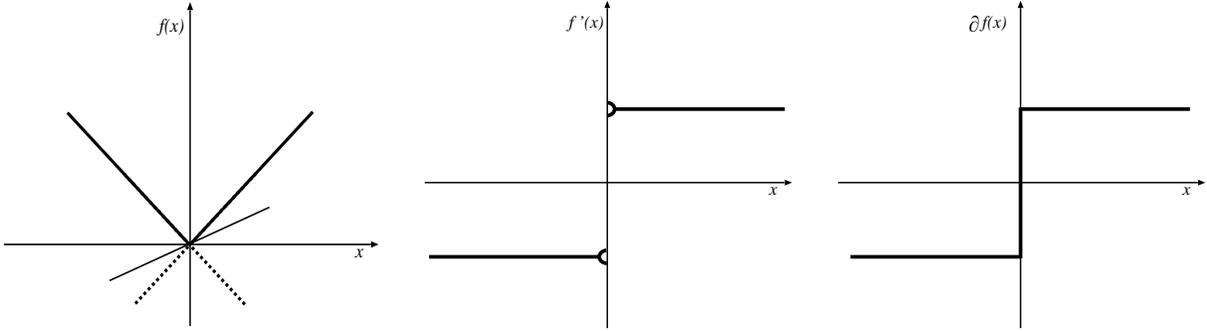}}
	\caption[A non-smooth function and its derivatives] {The geometrical representation of (a) the function $f(x)=|x|$, (b) its classical derivative $f'(x)$ and (c) the generalized derivative $\partial f(x)$.}
	\label{fig:f_derivs}
\end{figure}

If the function is piecewise continuous instead, having a non-smooth point at $x$, for example a kink as shown in Fig. (\ref{fig:f_derivs}), then the function $f(x)$ will not be absolutely differentiable at every point $x$. However such a function will possess at each point $x$, a left and a right derivative defined as, refer \cite{leine06b}
\begin{eqnarray}
f'_-(x) = \;\stackrel{limit}{_{y\uparrow x}}\dfrac{f(y)-f(x)}{y-x}, \nonumber \\
f'_+(x) =\; \stackrel{limit}{_{y\downarrow x}}\dfrac{f(y)-f(x)}{y-x}.
		\label{eqn:left_right_deriv}
\end{eqnarray}
Then the generalized derivative of the function $f(x)$ at the non-smooth point $x$ is declared as \textit{any value $f'_q(x)$} lying between its left and right derivatives \cite{clarke98}. Such an intermediate value can be expressed as a convex combination of the left and right derivatives,
\begin{equation}
f'_q(x) = (1-q)f'_-(x)+ qf'_+(x), \qquad 0 \leq q \leq 1.
		\label{{eqn:gen_deriv}}
\end{equation}
Geometrically , a generalized derivative is the slope of any line drawn through the point $(x,f(x))$ and between the left and right tangent lines. The set of all generalized derivatives of $f$ at $x$, or in general the convex hull of the extreme values of the derivative, is termed as the \textit{generalized differential} of $f$ at the point $x$,
\begin{eqnarray}
\partial f(x) & = & \overline{co} \left\lbrace  f'_-(x),f'_+(x)\right\rbrace \nonumber \\
				& = & \left\lbrace f'_q(x) | f'_q(x)= (1-q)f'_-(x)+ qf'_+(x), \; 0 \leq q \leq 1\right\rbrace .
		\label{eqn:gen_diff}
\end{eqnarray}
The generalized differential of Clarke at $x$ is then the set of the slopes of all the lines included in the cone bounded by the left and right tangent lines and is a closed convex set. It is defined as
\begin{equation}
\partial f(x) = \;\stackrel{\bigcap}{_{\delta > 0}} \overline{co}\left\lbrace 
\nabla f(y)|y \in x +B_\delta(0)\right\rbrace\;\subset \mathcal{R}^{n \times m},
		\label{eqn:clarke_gen_diff}
\end{equation}
with the gradient
\begin{equation}
\nabla \mathbf{f(x)} = \left( \dfrac{\partial \mathbf{f(x)}}{\partial \mathbf{x}}\right) ^T \; \in \mathcal{R}^{n \times m}.
		\label{eqn:clarke_gradient}
\end{equation}
The \textit{generalized jacobian of Clarke} is then defined as the transpose of the generalized differential, namely
\begin{equation}
\mathbf{J(x)}=(\partial \mathbf{f(x)})^T \;\subset \mathcal{R}^{n \times m}
		\label{eqn:clarke_jacobian}
\end{equation}

\emph{Bifurcations} refer to the qualitative changes in the structural behaviour of dynamical systems. In smooth dynamical systems which have smoothly varying vector fields, these bifurcations are associated with an eigenvalue or a pair of complex eigenvalues, passing through the imaginary axis under the variation of a control parameter. This automatically implies the dependency of the Jacobian matrices on the control parameters. However nonsmooth continuous systems possess discontinuity boundaries over which the vector fields are non-smooth. Due to this reason the classical Jacobian matrices cannot be obtained using the usual approach. It is in this context that people have applied the concept of Clarke's generalized differential to obtain the Jacobian matrix and its eigenvalues at the discontinuity boundaries. 

Let us consider an autonomous non-smooth continuous system described by Eq. (\ref{eqn:pw_deg}). Let $x_\mu$ be an equilibrium point of this system for some value of $\mu$. This means that the Jacobian matrix which defines locally the vector field around the equilibrium point is single valued and is defined by.
\begin{equation}
J(x_\mu, \mu) = \left \{
						\begin{array}{lcl}
							J_-(\mathbf{x}_\mu, \mu) & = & \left. \dfrac{\partial F_1(\mathbf{x},\mu)}{\partial x} \right \rvert_{\mathbf{x}=\mathbf{x}_\mu}, \; \mathbf{x}_\mu \in S_1  \\ 
							J_+\mathbf{x}x_\mu, \mu) & = &\left.  \dfrac{\partial F_2(\mathbf{x},\mu)}{\partial x} \right \rvert _{\mathbf{x}=\mathbf{x}_\mu},\; \mathbf{x}_\mu \in S_2 . 
						\end{array}
					\right.
		\label{eqn:pw_deg1}
\end{equation}
However the Jacobian matrices \emph{jump} at the discontinuity boundary $\Sigma_{i,j}$. The \emph{generalized Jacobian matrix} is therefore a convex combination of two matrices $J_-(\mathbf{x},\mu)$ and $J_+(\mathbf{x},\mu)$. It gives the set of all values which the \emph{generalized Jacobian} can take on the discontinity boundary $\Sigma_{i,j}$, namely
\begin{eqnarray}
J(\mathbf{x},\mu) & = & \overline{co} \left\lbrace  J_-(\mathbf{x},\mu),J_+(\mathbf{x},\mu)
				\right \rbrace  \nonumber \\
		 & = &  \left\lbrace (1-q)J_-(\mathbf{x},\mu)+ qJ_+(\mathbf{x},\mu), \;\forall \; q\in [0,1] \right\rbrace . 
		\label{eqn:clarke_Jacobian}
\end{eqnarray}
Thus the generalized Jacobian can be visualised to give a unique path of eigenvalues during the \emph{jump} as auxiliary variable $q$ is varied from $0$ to $1$.

\bibliographystyle{ws-ijbc}
\bibliography{Bibliography}
\end{document}